\documentclass[twocolumn,epjc3]{svjour3}
\usepackage{graphics}
\usepackage{multicol}
\usepackage{cuted}
\usepackage{flushend}
\usepackage{microtype}
\usepackage{bbm}
\usepackage{amsmath}
\usepackage{mathtools}
\usepackage{caption}
\usepackage{subcaption}
\usepackage{amssymb}
\usepackage{mathrsfs}       
\usepackage[pdftex]{graphicx}
\usepackage{pgfplots}
\pgfplotsset{width=10cm,compat=1.9}
\usepackage{xargs}       
\usepackage{wasysym}
\usepackage{engrec}
\usepackage{enumerate}
\usepackage{appendix}
\usepackage{empheq}
\usepackage{float}
\usepackage{stmaryrd}
\usepackage[parfill]{parskip}
\usepackage[pdftex,breaklinks]{hyperref}
\usepackage{titletoc}
\usepackage{makecell}
\usepackage{lscape}
\usepackage{algorithm}
\usepackage{algorithmicx}
\usepackage{tensor}
\usepackage{algpseudocode}
\usepackage{blkarray}
\usepackage{multirow}
\usepackage{bigstrut}
\usetikzlibrary{plotmarks}
\usepackage{comment}
\usepackage{soul}
\usepackage{cite}

\definecolor{redbb}{RGB}{228,30,43}

\newcommand{\ket}[1]{|#1\rangle}
\newcommand{\bra}[1]{\langle#1|}

\newcommand{\elma}[3]{\langle#1|#2|#3\rangle}

\newenvironment{customlegend}[1][]{%
    \begingroup
    \csname pgfplots@init@cleared@structures\endcsname
    \pgfplotsset{#1}%
}{%
    \csname pgfplots@createlegend\endcsname
    \endgroup
}%
\def\addlegendimage{\csname pgfplots@addlegendimage\endcsname}

\begin{document}
\title{Structure of $^{128,129,130}$Xe through multi-reference energy density functional calculations}

\author{Benjamin~Bally\thanksref{ad:esnt,ad:dft,em:bb} \and Giuliano~Giacalone\thanksref{ad:itp,em:gg}  \and 
Michael~Bender\thanksref{ad:ipnl,em:mb}} 

\thankstext{em:bb}{\email{benjamin.bally@cea.fr}}
\thankstext{em:gg}{\email{giacalone@thphys.uni-heidelberg.de}}
\thankstext{em:mb}{\email{bender@ip2i.in2p3.fr}}

\institute{
\label{ad:esnt}
ESNT, IRFU, CEA, Universit\'e Paris-Saclay, F-91191 Gif-sur-Yvette, France 
\and
\label{ad:dft}
Departamento de F\'isica Te\'orica, Universidad Aut\'onoma de Madrid, E-28049 Madrid, Spain
\and
\label{ad:itp}
Institut f\"ur Theoretische Physik, Universit\"at Heidelberg, Philosophenweg 16, D-69120 Heidelberg, Germany
\and
\label{ad:ipnl}
Universit{\'e} de Lyon, Institut de Physique des 2 Infinis de Lyon, IN2P3-CNRS-UCBL, 4 rue Enrico Fermi, F-69622 Villeurbanne, France
}

\date{Received: \today{} / Revised version: date}

\maketitle

%
%
\begin{abstract}
Recently, values for the Kumar quadrupole deformation parameters of the nucleus $^{130}$Xe have been computed from the results of a Coulomb excitation experiment, indicating that this xenon isotope has a prominent triaxial ground state. Within a different context, it was recently argued that the analysis of particle correlations in the final states of ultra-relativistic heavy-ion collisions performed at the Large Hadron Collider (LHC) points to a similar structure for the adjacent isotope, $^{129}$Xe.
In the present work, we report on state-of-the-art multi-reference energy density functional calculations that combine projection on proton and neutron number as well as angular momentum with shape mixing for the three isotopes $^{128,129,130}$Xe using the Skyrme-type pseudo-potential SLyMR1.
Exploring the triaxial degree of freedom, we demonstrate that the ground states of all three isotopes display a very pronounced triaxial structure. Moreover, comparison with experimental results shows that the calculations reproduce fairly well the low-energy excitation spectrum of the two even-mass isotopes. By contrast, the calculation of $^{129}$Xe reveals some deficiencies of the effective interaction.

\end{abstract}
%
%
\section{Introduction}
\label{sec:intro}

Atomic nuclei exhibit collective correlations that often can be interpreted in terms of deformed intrinsic shapes. Within such a picture, the nuclear radius or, alternatively, the nucleon one-body density is then expanded as a series of multipole deformations in which the leading-order terms, i.e. quadrupolar and octupolar deformations, are the most important ones to interpret and understand the phenomenology of low-lying nuclear states.
While not observable per se, a large body of evidence has been gathered over the last fifty years that indicates that picturing nuclei as objects having an intrinsic shape captures essential information about their structure \cite{BM98a}. The question remains how to connect the parameters of such shapes to quantities that can be experimentally measured on the one hand, and to operator matrix elements of nuclear models on the other hand. 
For example, in the rigid-rotor model the magnitude of quadrupole deformation of the states belonging to a given rotational band can be related to the $E2$ transition probabilities within this band \cite{BM98a,RS80a}.
From a more general perspective, Ripka proposed to use the quadrupole moment and square charge radius to compute the parameters of an equivalent axially symmetric ellipsoid and applied the method in the context of Hartree-Fock calculations \cite{Ripka68a}. Few years later, Kumar showed how to relate the parameters of a general triaxial ellipsoid to the sum rules of matrix elements of various products of quadrupole operators that can be evaluated experimentally \cite{Kumar72a}, e.g.~in Coulomb excitation experiments. 
While not free of ambiguities \cite{Poves20a}, the latter method has been used to evaluate information about pronounced triaxial shapes of several nuclei over the past decades \cite{Cline86a,Ayangeakaa16a,Ayangeakaa19a,Henderson19a,Zielinska21p}. One such recent experiment focused on the nucleus $^{130}$Xe and determined the parameters $\left[ \beta_k (0^+_1) =0.17(2),~\gamma_k(0^+_1) =23(5)^\circ \right]$ \cite{Morrison20a}, which indicates prominent triaxial deformation.

Within a completely different context, recent experimental \cite{STAR:2015mki,ALICE:2018lao,CMS:2019cyz,ATLAS:2019dct,STAR:2021mii,ATLAS:2022dov} and theoretical \cite{Goldschmidt:2015kpa,Giacalone:2017dud,Giacalone:2018apa,Giacalone20a,Giacalone:2020awm,Jia:2021tzt,Bally22a,Jia:2021qyu,Zhang:2021kxj,Xu:2021uar,Nijs:2021kvn,Zhao:2022uhl,Magdy:2022cvt} investigations have demonstrated that angular (azimuthal) correlations of particles in the final states of heavy-ion collisions performed at ultra-relativistic energies at large scale facilities, such as the LHC, contain fingerprints of nuclear ground-state deformations. Motivated by our proposal \cite{Bally22a}, the ATLAS Collaboration has in particular shown \cite{ATLAS:2022dov} that data collected in high-energy $^{129}$Xe + $^{129}$Xe collisions is consistent with a triaxially-deformed shape $(\beta \approx 0.20, \gamma \approx 30^\circ)$ for the $J^\pi = 1/2^{+}$ ground state of $^{129}$Xe. 

In this article, we investigate the structure of the latter nucleus and the two adjacent even-even Xe isotopes in more detail using the same state-of-the-art multi-reference energy density functional (MR-EDF) approach as in Ref.~\cite{Bally22a}. The MR-EDF framework \cite{Bender03a,Schunck19} is a powerful tool that can be applied to the description of the structure of deformed nuclei and has been used extensively over the years in nuclear physics \cite{Bender03a,Egido16a,Robledo18a}. In particular, the use of symmetry-restoration and configuration-mixing methods allows one to compute the low-energy spectrum, including the electromagnetic transition probabilities, of both even- and odd-mass nuclei \cite{Bender08a,Rodriguez10a,Bally14a,Borrajo15a}.

This article is organized as follows: In Sec.~\ref{sec:theory}, we present the essential elements of our theoretical framework. Then, in Sec.~\ref{sec:resu}, we report the results of our new calculations on $^{128,129,130}$Xe.
Finally, we discuss the conclusions and perspectives in Sec.~\ref{sec:conclu}.

%
%
\section{Theoretical Framework}
\label{sec:theory}

We recall here only the most important aspects of our methodology. More information on the technical details of our implementation can be found for example in Refs.~\cite{Bender08a,Bally21a,Bonche87a}.

\subsection{Multi-reference ansatz}
\label{sec:mredf}

Within our MR-EDF implementation, we build approximations to the nuclear eigenstates by considering the variational ansatz
\begin{equation}
\label{eq:ansatzGCM}
|\Psi^{JMNZ\pi}_{\sigma} \rangle=\sum_{qK} f^{JMNZ\pi}_{\sigma;qK} \hat{P}^{J}_{MK} \hat{P}^{N} \hat{P}^{Z} |\Phi^\pi (q)\rangle ,
\end{equation}
where $\Omega \equiv \left\{|\Phi^\pi (q)\rangle, q \right\}$ is a set of Bogoliubov quasi-particle vacua that are used as reference states to calculate the densities that enter the MR-EDF, $\hat{P}^{N}$ and $\hat{P}^{Z}$ are the projectors onto good number of neutrons and protons, respectively, and $\hat{P}^{J}_{MK}$ is the angular momentum projection operator \cite{RS80a,Bally21a}. For the sake of compact notation,  we write the set of quantum numbers with the label $\Gamma \equiv (JMNZ\pi)$. The subscript $\sigma$ labels the order of the states with same quantum numbers $\Gamma$ in the energy spectrum.
The weights $f^{\Gamma}_{\sigma;qK}$ are variational parameters that are obtained through the minimization of the total energy for the \textit{ansatz} of Eq.~(\ref{eq:ansatzGCM}) that leads to a generalized eigenvalue problem that in the context of MR-EDF methods is known as
the Hill-Wheeler-Griffin (HWG) equation~\cite{Hill53a,Griffin57a}
\begin{equation}
 \label{eq:hwg}
 \sum_{q'K'} \left( \mathcal{H}^{\Gamma}_{qK,q'K'} - E^{\Gamma}_{\sigma} \mathcal{N}^{\Gamma}_{qK,q'K'} \right) f^{\Gamma}_{\sigma;q'K'} = 0 \, ,
\end{equation}
where
\begin{subequations}
\begin{align}
\label{eq:HqKqpKp}
 \mathcal{H}^{\Gamma}_{qK,q'K'}
 & = \langle \Phi^\pi(q)| \hat{H} \, \hat{P}^{J}_{KK'}\, \hat{P}^{N} \, \hat{P}^{Z} |\Phi^\pi(q')\rangle \, , \\
 \label{eq:NqKqpKp}
 \mathcal{N}^{\Gamma}_{qK;q'K'}
 & = \langle \Phi^\pi(q)|\hat{P}^{J}_{KK'} \, \hat{P}^{N} \, \hat{P}^{Z}|\Phi^\pi(q')\rangle  \, ,
\end{align}
\end{subequations}
are the Hamiltonian and norm matrices. The appearance of the latter is a consequence of the projected reference states being in general non-orthogonal. Note that the mixing of $K$ components is the necessary final step of angular-momentum projection even when not mixing states projected from different reference states~\cite{Bally21a}.

Concerning their numerical parameters, we use $M_{N} = M_{Z} = 7$ points for the discretization of the particle-number projection operators and $M_{\alpha} \times M_{\beta} \times M_{\gamma} = 40 \times 48 \times 40$ points for the discretization of the angular-momentum projection operator, in both cases using the prescriptions described in Ref.~\cite{Bally21a}.

Solving the HWG equation gives access to the energies $E^{\Gamma}_{\sigma}$ of the correlated states $|\Psi^{\Gamma}_{\sigma} \rangle$ as well as the weights $f^{\Gamma}_{\sigma;qK}$ of the projected reference states that can then be used to compute any observable of interest. Traditionally, the HWG equation~\eqref{eq:hwg} is solved by first diagonalizing the norm matrix, which provides the so-called "collective subspace" of the calculation. The lowest norm eigenvectors corresponding to an eigenvalue smaller than a suitably chosen cutoff are then discarded. The latter correspond to states that cannot be numerically distinguished from redundant ones in the non-orthogonal basis, and have to be removed in order to avoid numerical problems in the following steps \cite{Bonche90a}. Finally, the Hamiltonian matrix is then diagonalized in the reduced subspace of orthogonal norm eigenstates.

Besides the already mentioned one, a few further cutoffs are applied along the way. Before the mixing, we remove the projected components that in the decomposition of the original reference states have a weight that is lower than $10^{-4}$ (except for the $J^\pi = 11/2^-$, $13/2^-$ and $15/2^-$  states of $^{129}$Xe for which we use $10^{-3}$ instead).  During the mixing of $K$-components, performed individually for each reference state, we remove the norm eigenstates with an eigenvalue smaller than $10^{-5}$ for all nuclei. Finally, during the final diagonalization mixing projected states originating from different Bogoliubov vacua, we remove the norm eigenstates with an eigenvalue smaller than $10^{-5}$ for all nuclei. 

The elements of the Hamiltonian matrix and all other operator matrix elements are computed using the Generalized Wick Theorem (GWT) \cite{Balian69a} expressed in the non-equivalent canonical bases of the two Bogoliubov quasi-particle vacua that enter their calculation \cite{Bonche90a}. Following an original idea by Robledo~\cite{Robledo09a}, the overlaps between Bogoliubov quasi-particle vacua that enter the expressions for the norm kernels and operator matrix elements are computed from the Pfaffian expression derived in Ref.~\cite{Avez12a} that is adapted to our numerical representation.


\subsection{Reference states}
\label{sec:bogo}

In the present article, the reference states making the set $\Omega$ are Bogoliubov quasi-particle vacua that are chosen to be eigenstates of parity and $x$-signature, and also to be invariant under a $y$-time-simplex transformation, see Ref.~\cite{Bally21a} for further details. The three associated operators, $\hat{P}$,  $\hat{R}_x$, and $\hat{S}^{\mathcal{T}}_y \equiv \hat{R}_y \hat{P} \hat{\mathcal{T}}$, with $\hat{\mathcal{T}}$ being the time-reversal operator and $\hat{R}_y$ the $y$-signature operator, generate a subgroup of the double point-symmetry group $D_{\text{2h}}^{\text{TD}}$ as defined in Refs.~\cite{Dobaczewski00a,Dobaczewski00b}. With this set of symmetries, one can explore triaxial deformations, perform cranking calculations and generate Bogoliubov vacua with either an even or odd number parity. In particular, the latter are used to describe odd-mass nuclei.

The different Bogoliubov quasi-particle vacua in the set $\Omega$ are generated by solving self-consistently the Hartree-Fock-Bogoliubov (HFB) equations subject to a set of constraints imposed during the minimization, which for the present calculations consists of the average numbers of protons $Z_q$ and  neutrons $N_q$, as well as two parameters $(\beta, \gamma$) that characterize quadrupole deformation in the intrinsic major-axis system and whose definition is specified below. All other multipole moments compatible with the chosen symmetries are left free to take the value that gives the lowest total energy for a given set of constraints. Calculating operator matrix elements with the GWT requires that the Bogoliubov quasi-particle vacua are all non-orthogonal. To avoid the collapse of pairing correlations when solving the HFB equations, which can possibly lead to orthogonal reference states, the average particle numbers are shifted compared to their physical values, i.e.\ we take $Z_q = Z + 0.2$ and $N_q = N + 0.2$ with $Z=54$ and $N=74$, 75, 76. The subsequent projection on particle number then restores the correct particle number for all operator matrix elements discussed in what follows. This choice is somewhat arbitrary and we did not try to optimize the value of the shift to obtain the lowest projected states possible. 

The Bogoliubov reference states are represented on a three-dimensional cartesian Lagrange mesh \cite{Baye86a} in a box of 28 points in each direction and spaced by 0.8 fm. Taking advantage of the point-group symmetries, it is sufficient to numerically solve the HFB equations in one eighth of the full box \cite{Ryssens15a}, and to calculate the projected matrix elements in one half.

To explore the deformation space in an efficient manner, we used in practice an equally-spaced mesh in the coordinates ($q_1, q_2$) that are related to the ($\beta,\gamma$) through~\cite{Bender08a}
\begin{subequations}
\begin{align}
q_1 & = Q_0 \, \cos(\gamma) - \frac{1}{\sqrt{3}} \, Q_0 \, \sin(\gamma) \, , \\
q_2 & = \frac{2}{\sqrt{3}} \, Q_0 \, \sin(\gamma)  \, . 
\end{align}
\end{subequations}
In these expressions, the moment $Q_0 = \sqrt{q_1^2 + q_2^2 + q_1 \, q_2}$ \\ $= \sqrt{\tfrac{2}{3} \big( Q_{xx}^2 + Q_{yy}^2 + Q_{zz}^2 \big) }$ with $\hat{Q}_{\mu\nu} \equiv 3 \, \hat{r}_\mu \hat{r}_\nu - \hat{r}^2 \, \delta_{\mu\nu}$, is the total cartesian quadrupole moment \cite{Ryssens15a} that is related to $\beta$ through
\begin{equation}
\beta
= \sqrt{\frac{5}{16 \pi}} \, \frac{4 \pi \, Q_0}{3 R_0^2 A}
\end{equation}
with $R_0 \equiv 1.2 \, A^{1/3} \, \text{fm}$. More precisely, we used a mesh with a spacing\footnote{When considering the axial axis, this corresponds to a spacing $\Delta \beta \approx 0.05$.} $\Delta q_1 = \Delta q_2 = 175$ fm$^2$ starting from $(q_1,q_2) = (0,0)$ and restricting ourselves to positive values of $q_1$ and $q_2$, which maps the first sextant of the $\beta$-$\gamma$ plane. 

In the present article, the sets of reference states employed to describe the even-mass isotopes $^{128,130}$Xe contain only time-reversal-invariant Bogoliubov vacua such that all sextant are strictly equivalent \cite{Bender08a,Rodriguez10a}.
By contrast, in order to describe the odd-mass nucleus $^{129}$Xe, it is necessary to use reference states that are non-invariant under the time-reversal operation and, therefore, lead to slightly different results in the different sextants due to the relative orientation of their angular momentum with respect to the signature axis \cite{Schunck10a}. We neglect this effect in the present study.

We also remark that in the energy surfaces showed below, we plot the projected energies as a function of the average deformation of the Bogoliubov quasi-particle vacua they are projected from.

In addition, it is important to mention that our framework assumes point-like particles. Therefore, when computing the values for the root-mean-square (rms) charge radii $r_\text{rms}$ of the three isotopes, we correct for the finite size of the protons and neutrons using the prescription \cite{JodonPHD}
\begin{equation}
  r^2_\text{rms} = \langle r^2_\text{P} \rangle +  r_p^2 + \frac{N}{Z} r_n^2
\end{equation}
where $\langle r^2_\text{P} \rangle$ is the mean-square radius of the point-proton distribution in the many-body wave function, $r^2_p = 0.769$ fm is the mean-square charge radius of the proton and $r^2_n = - 0.116$ fm is the mean-square charge radius of the neutron.

Finally, we specify that the calculations of the electromagnetic transitions and moments were carried out using the bare electric charges and $g$-factors.\footnote{See Sec.~\ref{sec:spectrum129} for more comments on the matter.}

\subsection{Effective interaction}
\label{sec:hamil}

For the nuclear Hamiltonian, we use the SLyMR1 parameterization of the phenomenological Skyrme-type pseudo-potential. Its main specificities are that the commonly-used density-dependent terms are replaced by the central three-body forces with up to two gradients that have been constructed in Ref.~\cite{Sadoudi13a}, and that the associated energy functional is constructed as the strict expectation value of a true operator keeping all direct, exchange, and pairing terms, such that the same interaction is used in the normal and pairing channels. Because of this, and unlike the vast majority of parameterizations of the nuclear EDF used in the literature, SLyMR1 can be safely used in MR-EDF calculations without any of the ambiguities outlined in Refs.~\cite{Dobaczewski07a,Lacroix09a,Bender09a,Duguet09a}. For the same reason, the exchange and pairing contributions from the Coulomb interaction to the Hamiltonian matrix, Eq.~\eqref{eq:HqKqpKp}, are also calculated exactly.\footnote{When generating the deformed quasi-particle vacua, however, we neglected Coulomb pairing and used the much less costly Slater approximation for the Coulomb exchange energy.}

The coupling constants of this parameterization were fitted to reproduce the binding energies and charge radii of selected doubly-magic nuclei as well as basic properties of nuclear matter \cite{JodonPHD}. To be usable without ambiguity in time-reversal-breaking calculations, SLyMR1 was adjusted such that in all spin-isospin channels it is free of finite-size instabilities at all densities encountered in finite nuclei, see also Ref.~\cite{Pastore15a}. In spite of having a slightly larger number of parameters, this form of EDF does not reach the same overall quality of standard density-dependent EDFs as some of its parameters turn out to be over-constrained. For our calculations, the most relevant shortcomings are the very low isoscalar effective mass of $m^*/m = 0.53$ that leads to a larger than usual spacing of single-particle levels, and the overall very weak pairing strength, that in combination with the low effective mass can lead to a breakdown of HFB pairing correlations at many deformations.

To avoid a divergence in the pairing channel of the (zero-range) interaction \cite{Bulgac02a}, when solving the HFB equations we use a pairing window of 8.5 MeV above and below the Fermi energy outside of which the pairing interaction is progressively cut off. The cutoff is then omitted when calculating the Hamiltonian matrix, Eq.~\eqref{eq:HqKqpKp}, in the MR calculation.

%
%
\section{Results}
\label{sec:resu}

\subsection{$^{128}$Xe}
\label{sec:128}

We start our calculations by generating a set of Bogoliubov quasi-particle vacua that differ by their average triaxial deformation as described in Sec.~\ref{sec:bogo}. All vacua constructed have a good parity $\pi = +1$ and are invariant under time-reversal operation, which greatly simplifies the subsequent projected calculations.

\subsubsection{Energy surfaces}
\label{sec:surface128}

\begin{figure}[t!]
    \centering
    \includegraphics[width=0.70\linewidth]{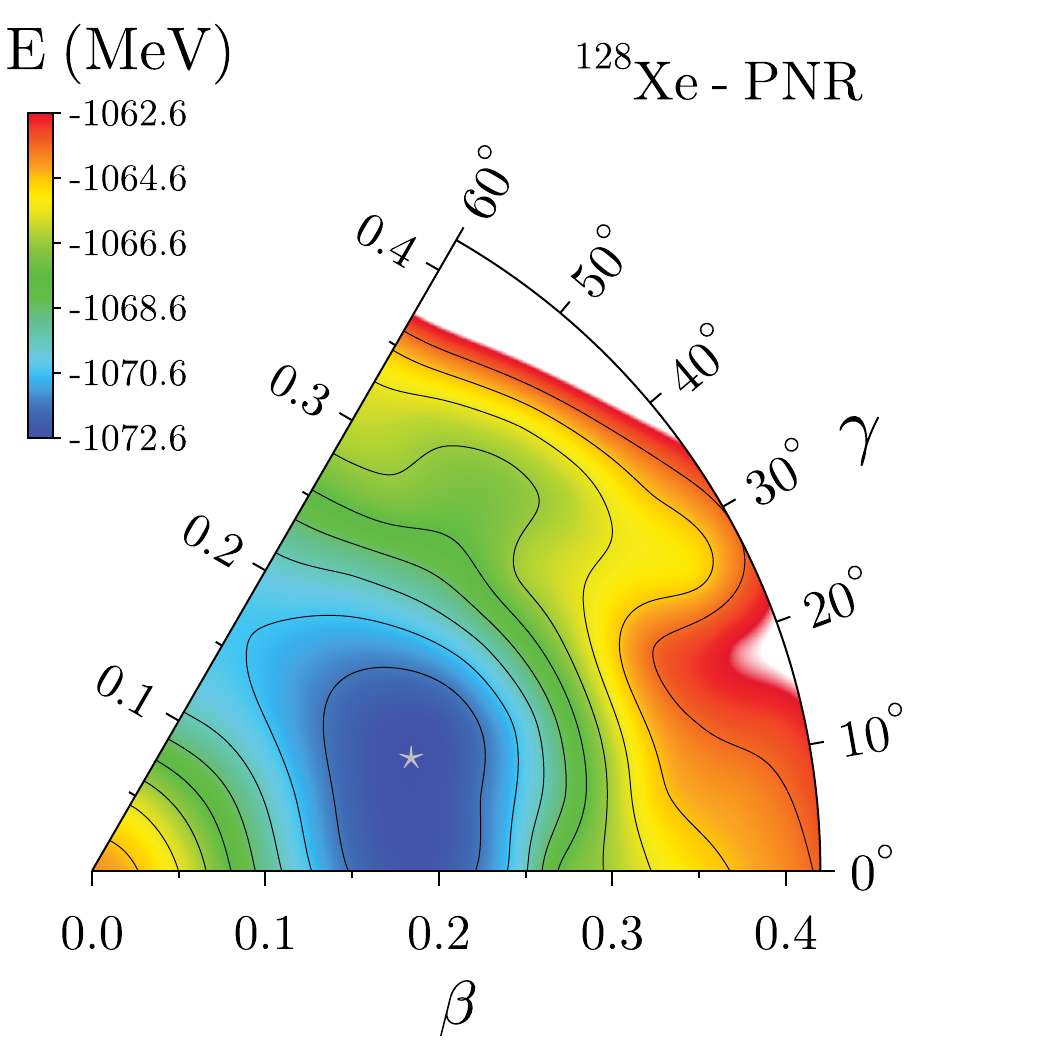} 
    \caption{Particle-number restored total energy surface for $^{128}$Xe. Black lines are separated by 1 MeV.
    The minimum, indicated by a silver star, is located at a deformation of $\beta=0.20$ and $\gamma=19^\circ$.}
     \label{fig:xe128_ener_pnr}
\end{figure}

\begin{figure}[t!]
    \centering
    \includegraphics[width=0.70\linewidth]{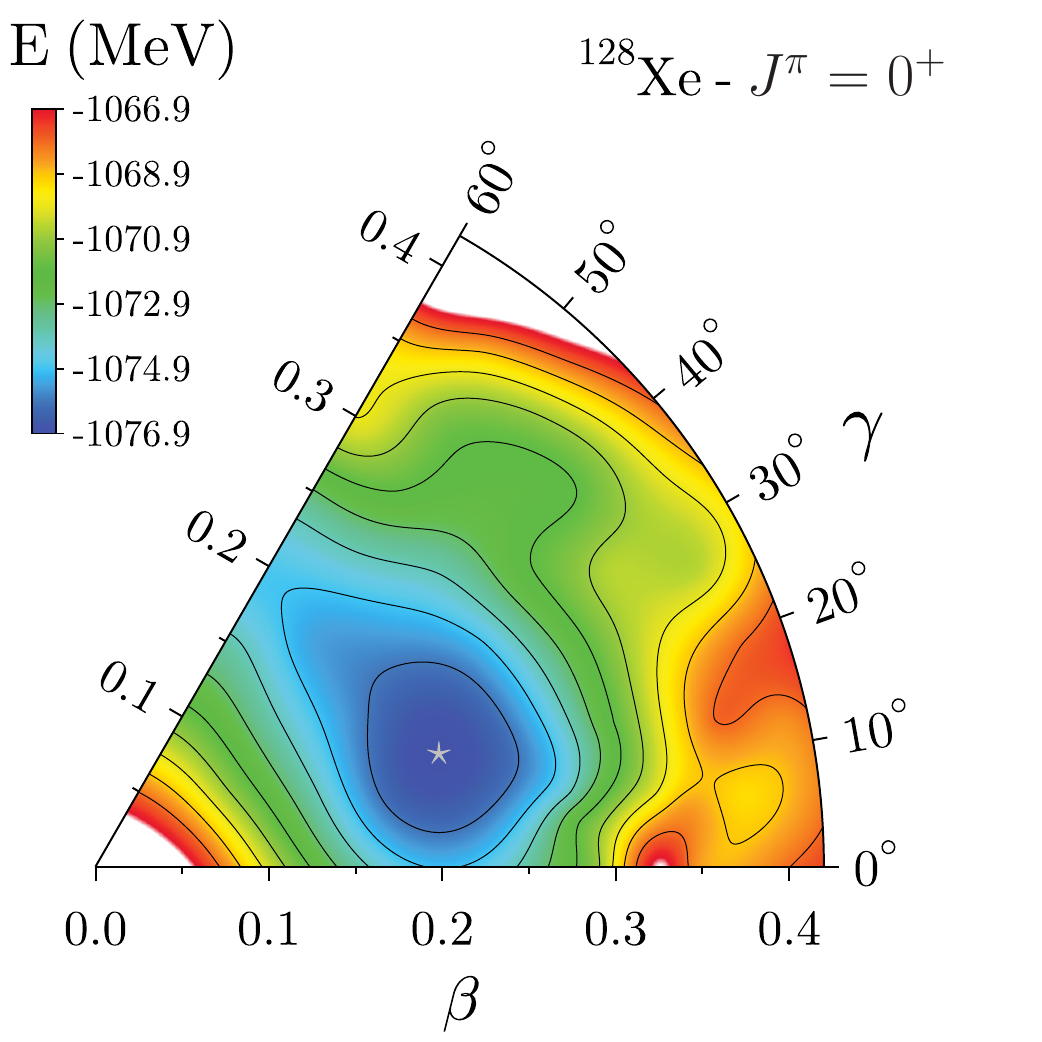} 
    \caption{Angular-momentum and particle-number restored total energy surface for $^{128}$Xe and $J^\pi = 0^+$. Black lines are separated by 1 MeV.
    The minimum, indicated by a silver star, is located at a deformation of $\beta=0.21$ and $\gamma=18^\circ$.}
    \label{fig:xe128_ener_ampnr}
\end{figure}

Figure~\ref{fig:xe128_ener_pnr} displays the particle-number restored (PNR) total energy surface for $^{128}$Xe. As can be seen, the minimum\footnote{Note that all the extrema discussed in this article are computed from an interpolation based on the results obtained at the points on the discretized deformation mesh as defined in Sec.~\ref{sec:bogo}.} of the energy surface is localised at pronounced triaxial deformation of $\beta=0.20$, $\gamma=19^\circ$, however, the surface is rather soft against $\gamma$-deformation.

Performing the full symmetry restoration, we display in Fig.~\ref{fig:xe128_ener_ampnr} the angular-momentum and particle-number restored total energy surface for the $J^\pi = 0^+$ projected out state at each deformation. We first observe that the minimum is lowered by about 4 MeV compared to a calculation were only particle-number is restored, but its deformation barely changes to the values $\beta=0.21$ and $\gamma=18^\circ$. The surface around the minimum, however, becomes more rigid in $\gamma$ direction and towards smaller values of $\beta$. Towards larger values of $\beta$, however, the surface becomes slightly softer. All of these changes are similar to what has been observed in previous MR-EDF calculations \cite{Bender08a,Rodriguez10a,Yao10a,Bally14a}, which all have shown that the energy gain from angular-momentum projection increases with deformation and tends to push minima towards more symmetry-breaking shapes.

\subsubsection{Low-energy spectroscopy}
\label{sec:spectrum128}

In order to select the reference states that are to be included in the set $\Omega$, we first sort all calculated vacua by the energy of their $J^\pi = 0^+$ component, and then keep all states with a projected energy up to 5~MeV above the lowest one. This leads to a set of 29 Bogoliubov quasi-particle vacua whose projected matrix elements enter the configuration mixing.

\begin{figure}[t!]
    \centering
    \includegraphics[width=.80\linewidth]{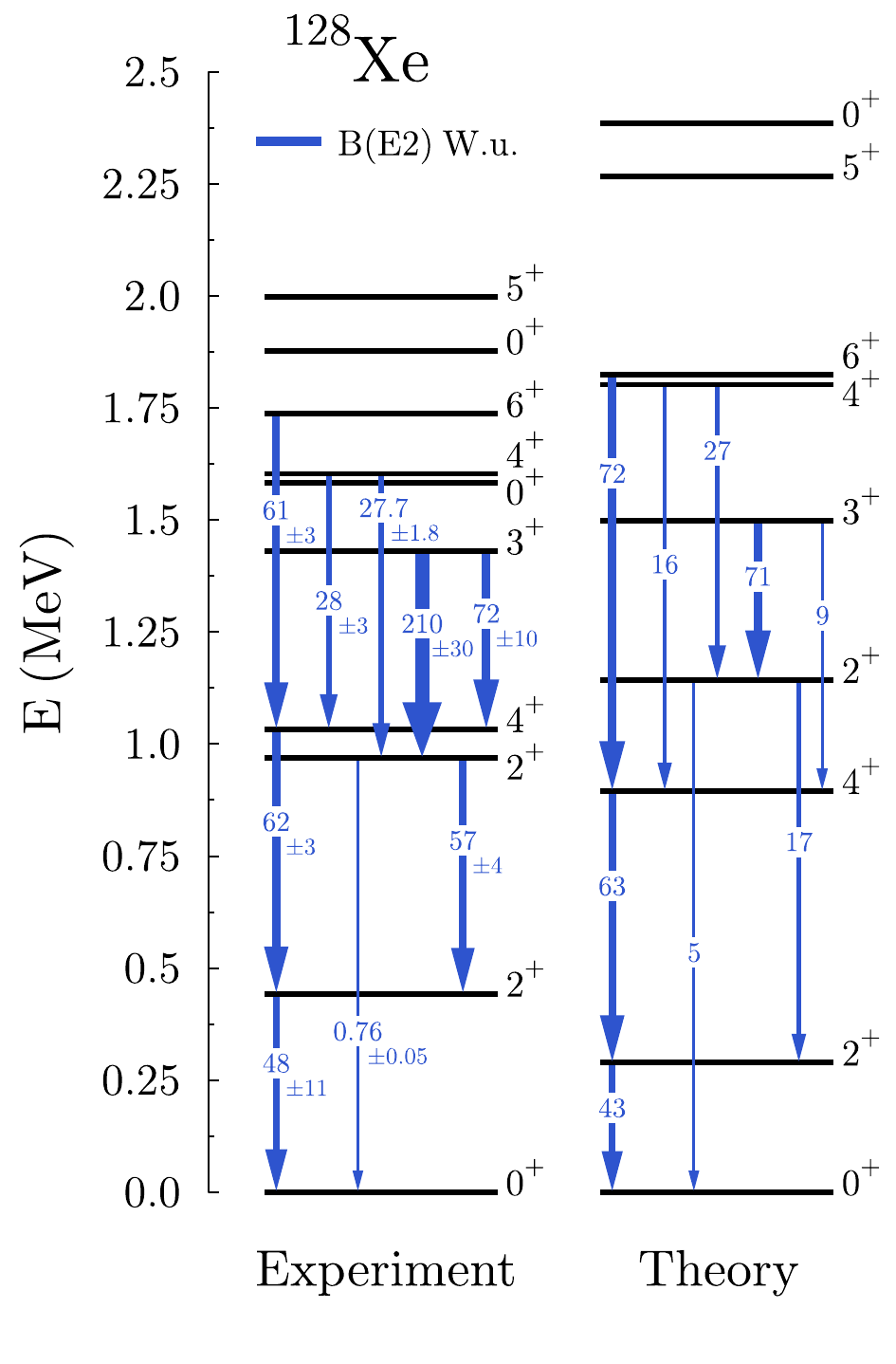}
    \caption{Low-energy spectrum for $^{128}$Xe. The $B(E2)$ transition probabilities are given in Weisskopf units (W.u). Experimental data are taken from \cite{AIEAa}, which are based on the evaluation \cite{Elekes15a}.}
    \label{fig:xe128_spec}
\end{figure}

In Fig.~\ref{fig:xe128_spec} we compare the theoretical results to the available experimental data for the low-lying excited states and the most relevant reduced $E2$ transition probabilities. 
The latter are given in Weisskopf units $5.940 \times 10^{-2} \, A^{4/3} \, e^2 \mathrm{fm}^4$,
which are an estimate for the typical contribution of one single-particle state to the transition probability \cite{RS80a}.
Our calculation also provides $B(M1)$ values, but they are usually too small by two orders of magnitude compared to experimental values and therefore are not shown. We will comment on possible reasons for this finding in Sec.~\ref{sec:spectrum129}.
Overall, we obtain a correct description of the spectrum (up to $\approx$ 2 MeV) with all the calculated levels appearing close to the experimental excitation energy except for the $0^+_2$ and $0^+_3$ that are too high. With 3.5 MeV, the calculated excitation energy of the latter is outside of the energy interval displayed on the plot. The other levels can be grouped into a $\Delta J = 2$ ground-state band and a $\Delta J = 1$ gamma-band, as expected for a nucleus of triaxial shape.
The $2^+_1$ and $4^+_1$ levels belonging to the ground-state band are slightly too low in energy, which possibly can be attributed to the overall too weak pairing strength of the SLyMR1 parameterization. As a result, the $4^+_1$ and $2^+_2$ are inverted in our calculation compared to the experimental data. At higher spin, however, the rotational bands are somewhat too spread out, which is a known feature of MR calculations of even-even nuclei in which the high-$J$ levels are projected out from a time-reversal invariant reference state \cite{Borrajo15a}. Nevertheless, the $B(E2)$ values for the transitions within the ground-state band, i.e.\ $6^+_1 \rightarrow 4^+_1 \rightarrow 2^+_1 \rightarrow 0^+_1$, are very well reproduced. On the contrary, the inter-band transitions, e.g.\ $4^+_2 \rightarrow 4^+_1$ or $3^+_1 \rightarrow 4^+_1$, tend to be underestimated.

\begin{table}[t!]
\centering
\begin{tabular}{ccc}
\hline\noalign{\smallskip}
Quantity & Experiment & Theory \\
\noalign{\smallskip}\hline\noalign{\smallskip}
$E(0^+_1)$ & $-1080.743$ & $-1077.956$ \\[0.08cm]
$r_\text{rms}(0^+_1)$ & 4.7774(50) & 4.738 \\[0.08cm]
$\mu(2^+_1)$ & $+0.68(7)$  & $+0.6$ \\[0.08cm]
$\mu(4^+_1)$ &  & $+1.2$ \\[0.08cm]
$\mu(2^+_2)$ &  & $+0.6$ \\[0.08cm]
$Q_s(2^+_1)$ &  & $-0.8$ \\[0.08cm]
$Q_s(4^+_1)$ &  & $-0.9$ \\[0.08cm]
$Q_s(2^+_2)$ &  & $+0.8$ \\[0.08cm]
$B(E2:2_1^+ \rightarrow 0_1^+)$ & 1839(421) & 1664 \\
\noalign{\smallskip}\hline
\end{tabular}
\caption{Spectroscopic quantities for several states ($J^\pi_\sigma$) of $^{128}$Xe: binding energy $E$ (MeV), root-mean-square charge radius $r_\text{rms}$ (fm), dipole magnetic moments  $\mu$ ($\mu_N$), quadrupole moments $Q_s$ ($e$b), reduced transition probability $B(E2:2_1^+ \rightarrow 0_1^+)$ ($e^2$fm$^4$). Experimental data are taken from \cite{Angeli13a,Stone16a,Stone20a,AIEAa,Elekes15a,Huang21a,Wang21a}. The experimental error on the binding energy is much smaller than the rounded value given here.}
\label{tab:values128}
\end{table}

In Table~\ref{tab:values128}, we report spectroscopic quantities for some of the low-lying states. Unfortunately, not much is known experimentally. Still, we see that we reproduce fairly well the binding energy and root-mean-square charge radius
of the ground state, with a relative accuracy below 1\%, as well as the dipole magnetic moment of the $2^+_1$ state. Here, it is worth recalling that the calculations were carried out using the bare electric charges and $g$-factors.

\subsubsection{Average deformation}
\label{sec:avgdef128}

As previously remarked upon, the structure of atomic nuclei is often interpreted in terms of simple intrinsic shapes. 
But the assignment of an intrinsic shape to a given nuclear state is not a unique or unambiguous procedure and there exist several non-equivalent methods to compute deformation parameters for said state starting from either experimental data or theoretical calculations. The difference between these methods is that they are based on different model assumptions about quadrupole collectivity.

In the rigid-rotor model, one can attribute an intrinsic deformation to the states belonging to a rotational band by considering the $E2$ transitions within the band. For the ground-state band of an even-even nucleus, one obtains \cite{RS80a}
\begin{align}
\label{eq:beta:fro:BE2}
 \beta_{r} (0_1^+) 
 & = \frac{4\pi\sqrt{5}}{3ZR_0^2} \sqrt{B(E2:2_1^+ \rightarrow  0_1^+)} \nonumber \\
 & = \frac{4\pi}{3ZR_0^2} \,
     | \langle 0_1^+ || \hat{E}_{2} || 2^+_1 \rangle |
 \, ,
\end{align}
where the reduced transition probability is expressed in units of $e^2$fm$^4$. This so-called $B(E2)$ value is proportional to the reduced matrix element of the electric quadrupole operator, see Appendix \ref{sec:appred} for more details, which is then converted via a model assumption into a shape.

With the definition of Eq.~\eqref{eq:beta:fro:BE2}, we obtain from experimental data the value\footnote{We note in passing that three different recent data compilations provide slightly different values for the $B(E2)$. The one we list in Table~\ref{tab:values128} is taken from \cite{AIEAa},
while one can also find $1634(32)$ 
\cite{Elekes15a} and $1596(76)$
\cite{Pritychenko16a} for the same 
$B(E2:2_1^+ \rightarrow  0_1^+) = \tfrac{1}{5} \, B(E2:0_1^+ \rightarrow  2_1^+)$ value, and which yield 
values for $\beta_r (0^+_1)$ that are $0.192(2)$ and $0.1885(46)$, respectively.} 
$\beta_r (0^+_1)$ = 0.20(2), while using the calculated transition probability we obtain $\beta_r (0^+_1)$ = 0.19. The two values are obviously in very good agreement, which is a mere consequence of the fact that the $B(E2)$ values are themselves very close.

We note that for an asymmetric rigid-rotor, the spectroscopic quadrupole moments of the first and second $2^+$ states are always of same size but have opposite sign~\cite{Davydov58a}, which is the case for the calculated values reported in Table.~\ref{tab:values128}.
In Davydov's variant of the rigid rotor model with rigid moments of inertia \cite{Davydov58a} one has $E(2^+_1)+E(2^+_2) = E(3^+_1)$ irrespective of the value of the triaxiality angle $\gamma$ as long as it is non-zero (in which case the rotor has no $2^+_2$ and $3^+_1$ states). With $E(2^+_1)+E(2^+_2) = 1.412 \, \text{MeV}$ and $E(3^+_1) = 1.430 \, \text{MeV}$ this relation is almost fulfilled for experimental data. For calculated values, the agreement is similar with $E(2^+_1)+E(2^+_2) = 1.432$ and $E(3^+_1) = 1.498$. In Davydov's model the triaxiality angle $\gamma$ also determines the ratio $E(2^+_2)/E(2^+_1)$ through the relation
\begin{equation}
\frac{E(2^+_2)}{E(2^+_1)} = \frac{1+\sqrt{1-\frac{8}{9}\sin^2(3\gamma_d)}}{1-\sqrt{1-\frac{8}{9}\sin^2(3\gamma_d)}} \, .
\end{equation}
The experimental value of $E(2^+_2)/E(2^+_1) = 2.19$ implies $\gamma_d (0_1^+) = 27^\circ$, while the calculated MR value $E(2^+_2)/E(2^+_1) = 3.95$ yields $\gamma_d (0_1^+) = 19^\circ$, which is quite close to the values for $\gamma$
as found for the minima of the energy surfaces of Figs.~\ref{fig:xe128_ener_pnr} and~\ref{fig:xe128_ener_ampnr} and as extracted from the collective wave function that will be discussed below. The latter indicates that the overall characteristics of the low-lying states in the spectrum from the MR calculation of $^{128}$Xe can be re-interpreted in terms of a rigid asymmetric rotor in spite of the MR states having a much richer structure. 


\begin{figure}[t!]
    \centering
    \includegraphics[width=0.70\linewidth]{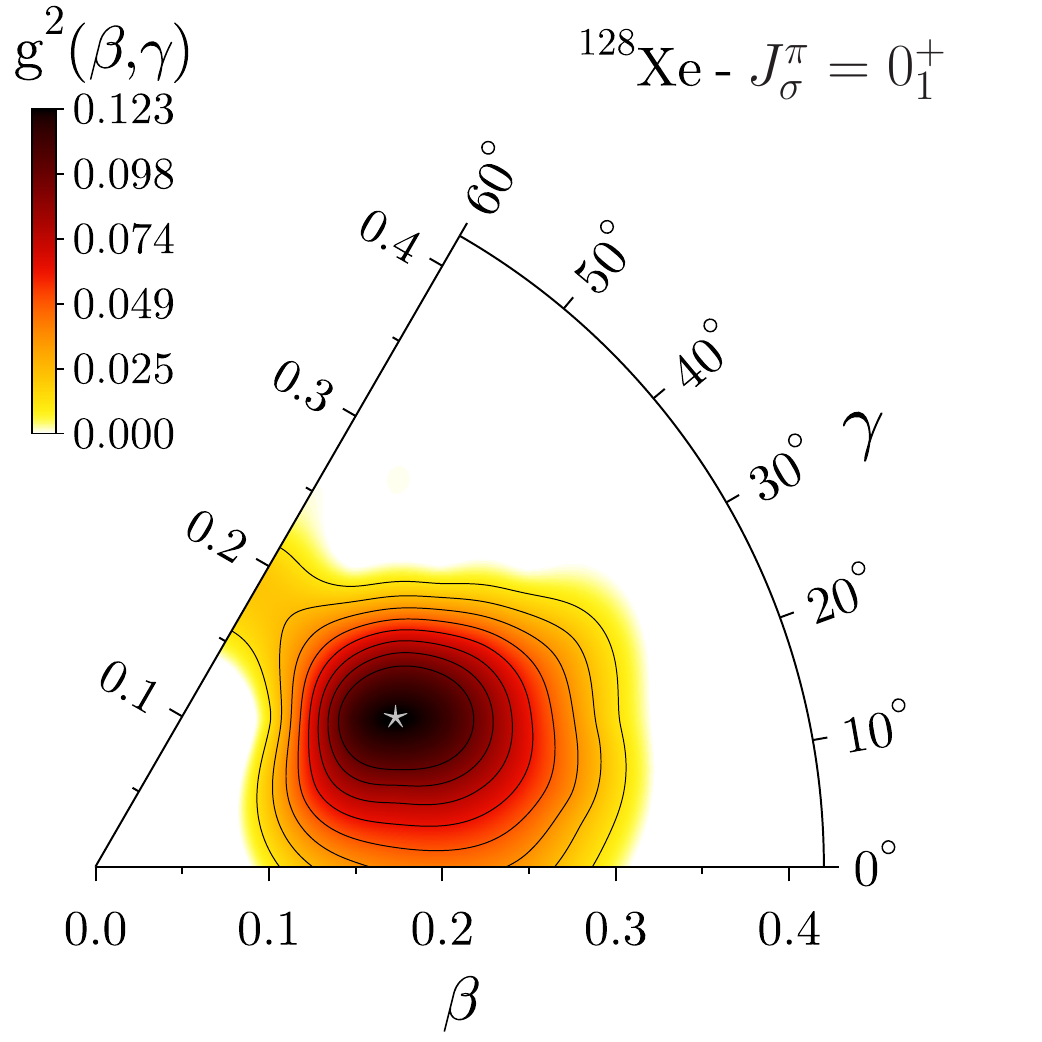}
    \caption{Collective wave function squared for the $0^+_1$ state of $^{128}$Xe. Black lines are separated by 10\% of the maximum value.
    The maximum, indicated by a silver star, is located at a deformation of $\beta=0.19$ and  $\gamma=26^\circ$.}
    \label{fig:xe128_coll}
\end{figure}

Another possibility to assign a deformation to the theoretical nuclear states is by looking at the decomposition of their correlated wave function in terms of the average deformation of the reference states included in set $\Omega$.
In Fig.~\ref{fig:xe128_coll}, we show the squared collective wave function (scwf) as a function of the triaxial deformation for the $0^+_1$ ground state. 
The collective wave function (cwf) $g^\Gamma_\sigma (q,K)$ for the components of Eq.~(\ref{eq:ansatzGCM}) is defined as
\begin{equation}
\label{eq:cwfdef}
g^\Gamma_\sigma (q,K) = \sum_{q'}^{\Omega} \sum_{K'=-J}^{J} \mathcal{N}^{\Gamma \: 1/2}_{qK;q'K'} \, f^{\Gamma}_{\sigma;q'K'},
\end{equation}
where $\mathcal{N}^{\Gamma \: 1/2}$ is the square-root matrix of $\mathcal{N}^{\Gamma}$, and satisfies the normalization condition\footnote{We remark that the surface obtained from the  interpolation of the cwf and displayed in Fig.~\ref{fig:xe128_coll} has not been normalized.} 
\begin{equation}
\label{eq:cwfnorm}
 \sum_{q}^\Omega \sum_{K=-J}^{J} \left[ g^\Gamma_\sigma (q,K)\right]^2 = 1 .
\end{equation}
We can regroup the square values of all components that have same value of $q$ but different values of $K$ by defining 
\begin{equation}
\label{eq:cwfgroup}
\left[ g^\Gamma_\sigma (q) \right]^2 \equiv \sum_{K=-J}^{J} \left[ g^\Gamma_\sigma (q,K)\right]^2 .
\end{equation}
The quantity $\left[ g^\Gamma_\sigma (q) \right]^2$ gives information about the spread of the correlated wave function as a function of the deformation, although strictly speaking it cannot be interpreted as a probability distribution due to the non-orthogonality of the reference states in the set.

Looking at Fig.~\ref{fig:xe128_coll}, we see that the ground state is peaked at triaxial shapes, as it could have been expected from the total energy surfaces presented in Figs.~\ref{fig:xe128_ener_pnr} and~\ref{fig:xe128_ener_ampnr}, with a maximum located at a deformation of $\beta=0.19$ and $\gamma=26^\circ$. But the scwf spreads in the triaxial plane with non-zero values ranging from $\beta \approx 0.1$ to $0.3$, while $\gamma$ covers a wide range of angles between the prolate axis and the oblate one. The fall-off of the scwf is, however, not the same in all directions, such that the average deformations found will not necessarily be equal to those at the maximum.

One possibility to quantify the role of fluctuations in the collective wave function is 
by making the rough approximation that the probability of finding a given intrinsic deformation is provided by the square of the scwf. Then, one can easily compute the moments of $\beta$ and $\gamma$
\begin{subequations}
\begin{align}
 \beta^{(n)}({\Gamma}_{\sigma}) &= \sum_{\beta,\gamma} \beta^n \, \left[ g^{\Gamma}_{\sigma} (\beta,\gamma) \right]^2 , \\
 \gamma^{(n)}({\Gamma}_{\sigma}) &= \sum_{\beta,\gamma} \gamma^n \, \left[ g^{\Gamma}_{\sigma} (\beta,\gamma) \right]^2 .
\end{align}
\end{subequations}
In particular, using the first- and second-order moments we can define the average deformations\footnote{A safer procedure to compute the average of an angle would be to first to compute the average of the complex number $\beta \exp(i\gamma)$ and then extract its argument. In practice, however, the two procedures give similar results because we explore here only the first sextant of the $\beta$-$\gamma$ plane.}
\begin{subequations}
\begin{align}
  \beta_{c}({\Gamma}_{\sigma}) &= \beta^{(1)}({\Gamma}_{\sigma}) , \\
  \gamma_{c}({\Gamma}_{\sigma}) &= \gamma^{(1)}({\Gamma}_{\sigma}) ,
\end{align}
\end{subequations}
as well as the corresponding standard deviations
\begin{subequations}
\begin{align}
  \Delta \beta_c({\Gamma}_{\sigma}) &= \left\{ \beta^{(2)}({\Gamma}_{\sigma}) -  \left[ \beta^{(1)}({\Gamma}_{\sigma})\right]^2 \right\}^{1/2} , \\
  \Delta \gamma_c ({\Gamma}_{\sigma}) &= \left\{ \gamma^{(2)}({\Gamma}_{\sigma}) -  \left[ \gamma^{(1)}({\Gamma}_{\sigma})\right]^2 \right\}^{1/2} ,
\end{align}
\end{subequations}
The interpretation of these quantities, however, has its limits as they are a statistical average and not a quantum-mechanical one. Their calculation also neglects the fact that we deal with a non-orthogonal set of reference states. Moreover, the fluctuations in $\beta$ and $\gamma$ are not independent, such that one should also analyse their covariance. Nevertheless, this simple procedure allows us to give a qualitative estimate of deformation fluctuations in the scwf.
When calculated for $^{128}$Xe, we obtain for the ground state an average elongation $\beta_{c}(0^+_{1}) = 0.22$, with a standard deviation $\Delta \beta_c(0^+_{1})= 0.05$, and an average angle $\gamma_{c}(0^+_{1}) = 21^\circ$,  with a standard deviation $\Delta \gamma_c(0^+_{1})= 12^\circ$. These values are consistent with the picture discussed above.

Finally, another popular method to associate a deformation to a nuclear state is through the calculation of the so-called Kumar quadrupole parameters \cite{Kumar72a,Cline86a,Srebrny06a} for an ``equivalent'' ellipsoid. The procedure goes as follows. 
First, one considers an ellipsoid with average electric quadrupole moments
\begin{subequations}
\begin{align}
  \langle E_{20} \rangle_v &= Q_k \cos(\gamma_k) , \\
  \langle E_{2\pm1} \rangle_v &= 0 , \\
  \langle E_{2\pm2} \rangle_v &= \frac{1}{\sqrt{2}}  Q_k \sin(\gamma_k) ,
\end{align}
\end{subequations}
where $\langle E_{2\mu} \rangle_v$ involves an integral over the volume of the ellipsoid. Like what is done to constrain the deformation of the quasi-particle vacua $q_1$ and $q_2$ or, equivalently, $\beta$ and $\gamma$, entering the MR calculation, the deformation of this ellipsoid is described by \emph{volume} parameters rather than $\emph{surface}$ parameters as often used, for example, in the context of the liquid drop model.
Then, in the next step one constructs powers of the electric quadrupole operator $\hat{E}_2$, which is a tensor of rank 2, coupled to zero total angular momentum, i.e.\ one considers products of the form
\begin{equation}
 \label{eq:e2n}
 \hat{E}_2^{(n)} \equiv \big[ \underbrace{\hat{E}_2 \times \ldots \times \hat{E}_2}_{n \text{ times}} \big]_0 \, ,
\end{equation}
where we remark that the decomposition in $\hat{E}_2^{(n)}$ is not necessarily unique as for high-order combinations there are several possibilities for intermediate coupling of these rank-2 tensor operators. 
Finally, the parameters of the ellipsoid are obtained by equating the expectation value of the operators $\hat{E}_2^{(n)}$ for a given nuclear state, $\ket{\Psi^\Gamma_\sigma}$, with the expanded right-hand side of Eq.~(\ref{eq:e2n}) where the operators $\hat{E}_{2\mu}$ have been replaced by the expectation values $\langle E_{2\mu} \rangle_v$, which is an approximation as one could in principle evaluate the expectation value $\langle E_{2}^{(n)} \rangle_v$ for the ellipsoid.

The second order product of the electric quadrupole operator allows one to compute the value of the elongation of the ellipsoid
\begin{align}
 \elma{\Psi^\Gamma_\sigma}{\hat{E}^{(2)}_2}{\Psi^\Gamma_\sigma} 
   &= \elma{\Psi^\Gamma_\sigma}{\left[ \hat{E}_2 \times \hat{E}_2 \right]_0}{\Psi^\Gamma_\sigma} \\
   &= \frac{1}{\sqrt{5}} Q_k^2 (\Gamma_\sigma) , \nonumber \\                
  \beta_k (\Gamma_\sigma) &\equiv \left( \frac{4\pi}{3R_0^2A} \right) \left[ \sqrt{5} \elma{\Psi^\Gamma_\sigma}{\hat{E}^{(2)}_2}{\Psi^\Gamma_\sigma} \right]^{1/2} ,
\end{align}
Computing the third order product of the electric quadru\-pole operator gives access to the triaxiality angle of the ellipsoid \begin{align}
 \elma{\Psi^\Gamma_\sigma}{\hat{E}^{(3)}_2}{\Psi^\Gamma_\sigma} 
   &= \elma{\Psi^\Gamma_\sigma}{ \left[ \left[ \hat{E}_2 \times \hat{E}_2 \right]_2 \times \hat{E}_2 \right]_0 }{\Psi^\Gamma_\sigma} \\
   &= -\sqrt{\frac{2}{35}}  Q_k^3 (\Gamma_\sigma) \cos\left[ 3\gamma_k (\Gamma_\sigma) \right]  , \nonumber \\
   \cos\left[ 3\gamma_k (\Gamma_\sigma) \right] &\equiv -\sqrt{\frac{35}{2}} \frac{\elma{\Psi^\Gamma_\sigma}{\hat{E}^{(3)}_2}{\Psi^\Gamma_\sigma}}{\left[ \sqrt{5} \elma{\Psi^\Gamma_\sigma}{\hat{E}^{(2)}_2}{\Psi^\Gamma_\sigma} \right]^{3/2}} .
\end{align}
Higher order products can be used to evaluate the moments of $\beta_k (\Gamma_\sigma)$ and $\gamma_k (\Gamma_\sigma)$. For example, it is necessary to compute $\elma{\Psi^\Gamma_\sigma}{\hat{E}^{(4)}_2}{\Psi^\Gamma_\sigma}$ to evaluate the standard deviation $\Delta \beta_k (\Gamma_\sigma)$ and to go up to the sixth order to compute the standard deviation  $\Delta \gamma_k (\Gamma_\sigma)$. Here, we use the particular prescription introduced in Ref.~\cite{Poves20a} to compute these fluctuations.

In experiment, however, one only has access to the matrix elements of $\hat{E}_2$.
Therefore, a crucial aspect of this procedure is that, by inserting the completeness relation for the Hamiltonian eigenstates between each subproducts of $\hat{E}_2$ operators coupled to zero angular momentum in Eq.~\eqref{eq:e2n} and by using the properties of $SU(2)$ irreducible tensor operators \cite{Varshalovich88a},  the matrix elements $\elma{\Psi^\Gamma_\sigma}{\hat{E}^{(n)}_2}{\Psi^\Gamma_\sigma}$ can be expressed as sums of products of reduced matrix elements of the type $\elma{\Psi^\Gamma_\sigma}{|\hat{E}_2|}{\Psi^{\Gamma'}_{\sigma'}}$ that can be both computed theoretically\footnote{In a MR-EDF approach, one could in principle evaluate the expectation values of $\hat{E}_2^{(n)}$ directly. Beginning with $n=3$, however, their calculation will become orders of magnitude more costly than the calculation of anything else in our implementation.} and measured experimentally, for example in Coulomb excitation experiments. The reduced expressions for $\hat{E}^{(n)}_2$, up to $n=6$, are given in Appendix~\ref{sec:appred}. In that sense, the matrix elements $\elma{\Psi^\Gamma_\sigma}{\hat{E}^{(n)}_2}{\Psi^\Gamma_\sigma}$ can be linked straightforwardly to experimental data and a meaningful comparison between theory and experiment is possible. Nevertheless, it is important to understand that associating these matrix elements to deformation parameters requires the additional assumption of an intrinsic state of classical ellipsoidal shape and is, therefore, entirely model dependent. 

\begin{table}[t!]
\centering
\begin{tabular}{ccc}
\hline\noalign{\smallskip}
Quantity & Theory & Experiment \\
\noalign{\smallskip}\hline\noalign{\smallskip}
$\beta_r$ & 0.19 & 0.20(2) \\[0.08cm]
$\gamma_d$ & $19^\circ$ & $27^\circ$ \\
\noalign{\smallskip}\hline\noalign{\smallskip}
$\beta_{c}$ & 0.22 & \\[0.08cm]
$\Delta \beta_c$ & 0.05 &  \\[0.08cm]
$\gamma_{c}$ & $21^\circ$ &  \\[0.08cm]
$\Delta \gamma_c$ & $13^\circ$ &  \\
\noalign{\smallskip}\hline\noalign{\smallskip}
$\beta_k$ & 0.20 &  \\[0.08cm]
$\Delta \beta_k$ & 0.02 &  \\[0.08cm]
$\gamma_k$ & $39^\circ$ &  \\[0.08cm]
$\Delta \gamma_k$ & $21^\circ$ &  \\[0.08cm]
\noalign{\smallskip}\hline
\end{tabular}
\caption{Deformation parameters for the $0_1^+$ ground state of $^{128}$Xe obtained by using the different methods detailed in the core of the text. The experimental values are computed using the energies and $B(E2)$ value taken from \cite{AIEAa,Elekes15a}.}
\label{tab:def128}
\end{table}

That being said, computing the Kumar quadrupole parameters both experimentally and theoretically can be an informative exercise as one should obtain similar deformation parameters if one starts with similar reduced matrix elements. In our case, for the ground states we obtain an elongation $\beta_k (0^+_1) = 0.20$, with a very small standard deviation $\Delta \beta_k (0^+_1) = 0.02$, and an angle $\gamma_k (0^+_1) = 39^\circ$, with a large standard deviation $\Delta \gamma_k (0^+_1) = 21^\circ$. 
Concerning the large value of $\Delta \gamma_k (0^+_1)$, we remark that in Ref.~\cite{Poves20a}, the authors argued that the fluctuations of the Kumar parameters are often large, in particular for $\gamma_k (\Gamma_\sigma)$, such that the interpretation of nuclear states in terms of well defined shapes has its limits.

In order to obtain a more general picture of the situation, in Table \ref{tab:def128} we compare the values of the average deformation parameters as given by the different methods. On the one hand, we see that the values of $\beta_r(0^+_{1})$, $\beta_{c}(0^+_{1})$ and $\beta_k (0^+_1)$ are very similar to each other and point towards an elongation of about 0.2.  On the other hand, there is a large discrepancy between the angles $\gamma_c (0^+_1)$ and $\gamma_{k}(0^+_{1})$ as well as the standard deviation $\Delta \gamma_c (0^+_1)$ and $\Delta \gamma_{k}(0^+_{1})$. With the values $\gamma_{k}(0^+_{1}) = 39^\circ$ and $\Delta \gamma_k (0^+_1) = 21^\circ$, Kumar's method suggests a triaxial shape centered in the upper half of the $0\textrm{--}~60^\circ$ sextant with rather large fluctuations. By contrast, the method based on the cwf indicates an average shape located in the first half of the sextant, $\gamma_c (0^+_1) = 21^\circ$, with somewhat smaller fluctuations, $\Delta \gamma_c (0^+_1) = 13^\circ$.

Ultimately, if one can hope that the two principal methods discussed here to compute an average deformation for a nuclear state, $\ket{\Psi^\Gamma_\sigma}$, give similar results, there is no guarantee that it will be the case. Indeed, the model-dependent definitions of the two sets of parameters are based on very different theoretical considerations. Chiefly, the deformation parameters defined using the cwf depend only on the correlated wave function of the state considered, whereas the sum rules appearing in the computation of the Kumar parameters involve reduced matrix elements between many intermediate states. More precisely, using the expressions of the matrix elements given in Appendix \ref{sec:appred}, we notice that while the calculations of $\beta_k (\Gamma_\sigma)$ and $\Delta \beta_k (\Gamma_\sigma)$ only make use of reduced matrix elements directly connected to  $\ket{\Psi^\Gamma_\sigma}$, the calculations of $\gamma_k (\Gamma_\sigma)$ and $\Delta \gamma_k (\Gamma_\sigma)$ involve also matrix elements between two different intermediate states. Given that the transition to the $0_1^+$ ground state is dominated by a single matrix element, $B(E2: 2_1^+ \rightarrow 0_1^+$), related to the decay of the $2_1^+$ rotational excitation, it explains why the estimates for $\beta_r(0^+_{1})$, $\beta_{c}(0^+_{1})$ and $\beta_k (0^+_1)$ agree so well. On the contrary, the too small value for the matrix element $B(E2: 2_2^+ \rightarrow 2_1^+$) probably prevents a good evaluation of the triaxiality angle, which would explain the surprising value for $\gamma_{k}(0^+_{1})$. In general, nuclear models do not reproduce equally well all the low-lying nuclear states and the transitions among them. Therefore, the usefulness of Kumar's procedure is sometimes limited even when the state of interest is well described.

That being said, all the deformation parameters calculated here as well as the energy surfaces and the scwf indicate that the ground state $^{128}$Xe can be understood in terms of an intrinsic quadrupole shape with an elongation of about 0.2 and a triaxial angle in the range $20\textrm{--}\,40^\circ$.
Concerning this large degree of triaxiality, we conclude by mentioning that within the Interacting Boson Model (IBM), the low-energy spectrum of $^{128}$Xe is well described by the $O(6)$ symmetry \cite{Arima79a}, which corresponds to $\gamma$-soft nuclei \cite{MeyerTerVehn79a}, but with deviations from the $O(6)$ limit case that suggest a nucleus having a more rigid triaxial deformation \cite{Casten85a}.

\subsection{$^{129}$Xe}
\label{sec:129}

\begin{figure}[t!]
    \centering
    \includegraphics[width=0.70\linewidth]{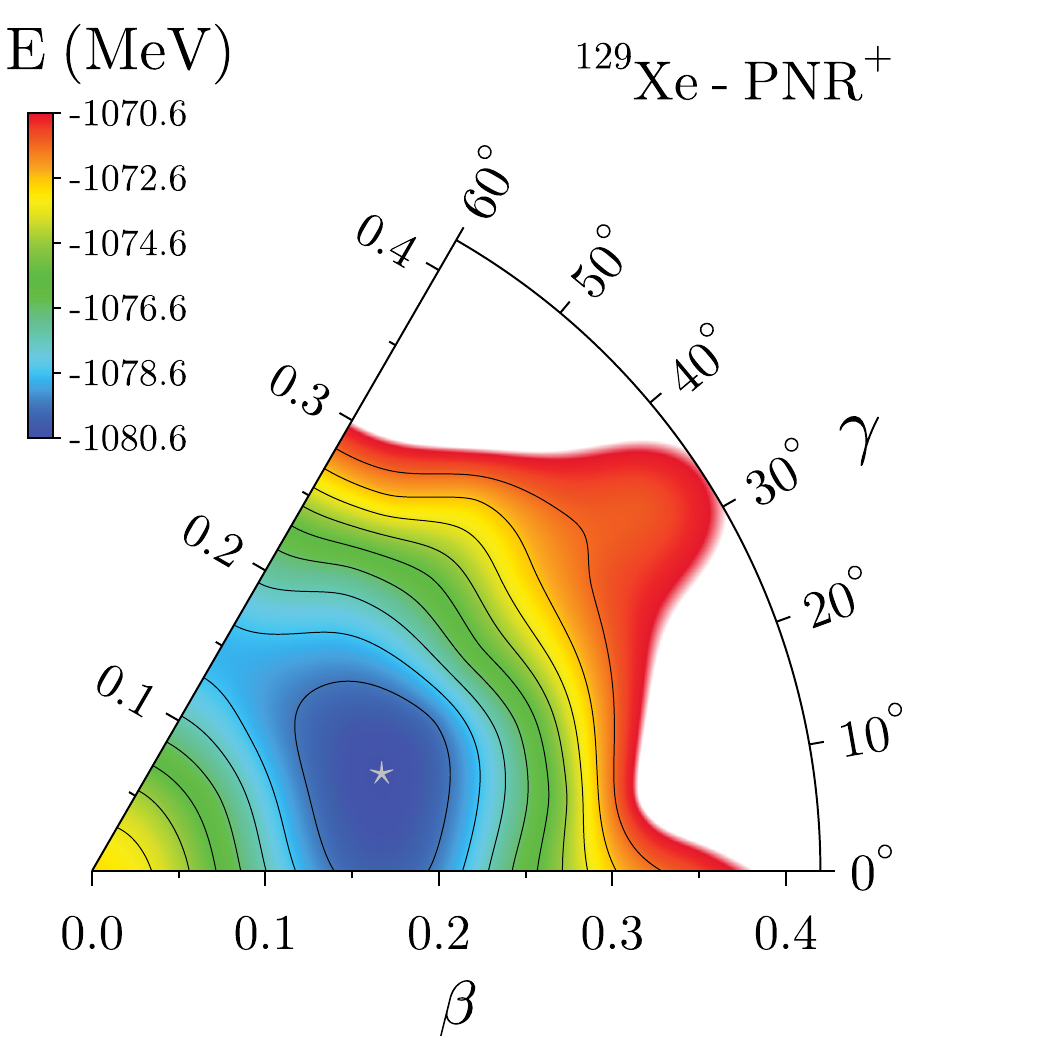} 
    \includegraphics[width=0.70\linewidth]{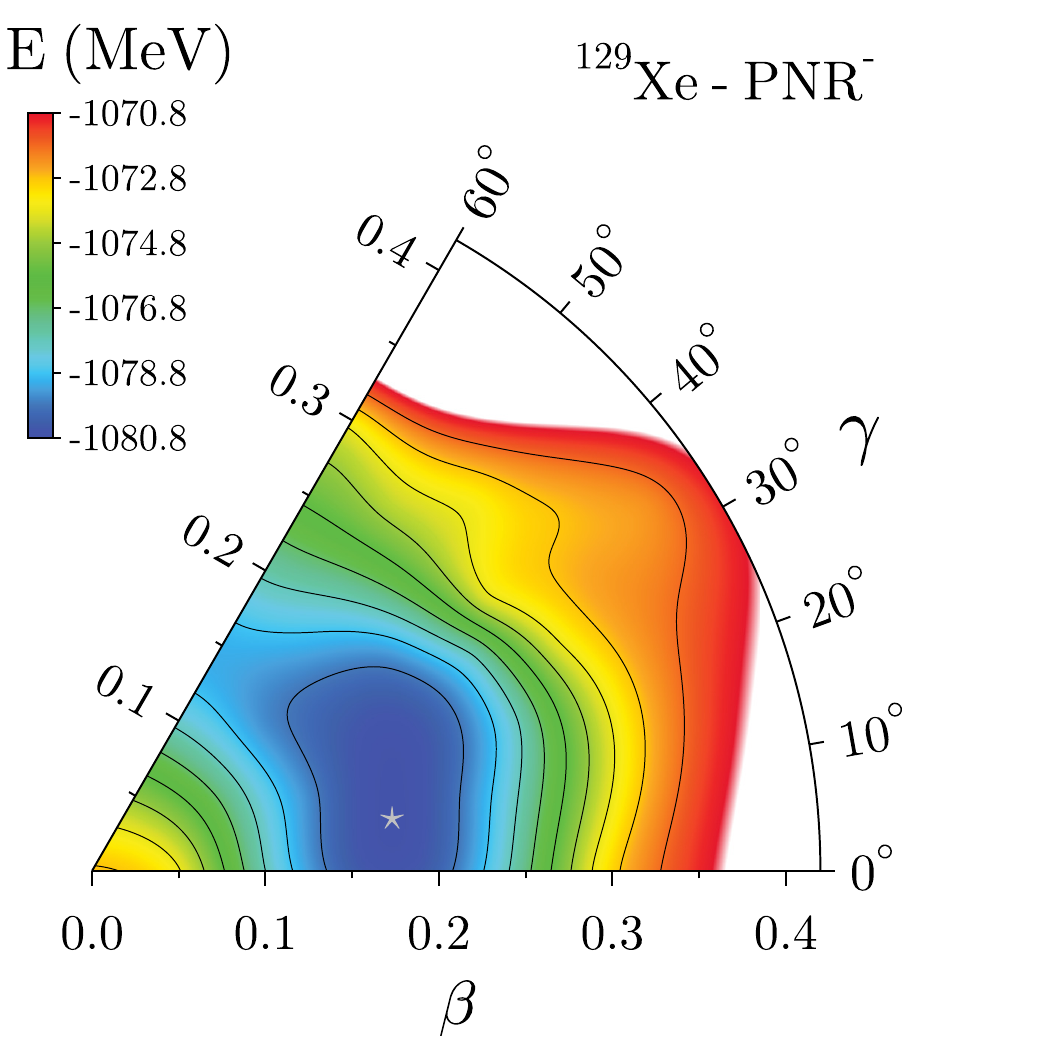}  
        \caption{Same as Fig.~\ref{fig:xe128_ener_pnr} but for $^{129}$Xe with $\pi = +1$ (top panel) and $\pi = -1$ (bottom panel). 
    The minimum for positive (negative) parity, indicated by a silver star, is located at a deformation of $\beta=0.18$ and $\gamma=19^\circ$ ($\beta=0.18$ and $\gamma=10^\circ$).}
     \label{fig:xe129_ener_pnr}
\end{figure}

The second nucleus we study is the odd-mass isotope $^{129}$Xe. In the past four years, the structure of this nucleus has been extensively discussed in the context of nuclear collisions at ultra-relativistic energies, due to the fact that a short run of $^{129}$Xe+$^{129}$Xe collisions has been performed at the LHC at the end of 2017. Based on the method introduced in Ref.~\cite{Giacalone20a}, we recently argued that not only fingerprints of the triaxial structure of its $1/2_1^+$ ground state are observed in ultra-relativistic collisions performed at the LHC \cite{ATLAS:2019dct}, but also that the magnitude of the quadrupole deformation parameters $(\beta,\gamma)$ of the associated intrinsic shape can be determined from the data \cite{Bally22a}. This has been subsequently demonstrated by the ATLAS Collaboration in a dedicated analysis \cite{ATLAS:2022dov}. Here, we realize a more thorough study of  this nucleus, and also compare our results with the available low-energy observables.

That low-lying states in $^{129}$Xe are dominated by triaxial shapes was inferred early on from the structure of its excitation spectrum \cite{Meyer76a,Helppi81a,Irving79a,Lonnroth83a,Huang16a,Zhao88a} that suggests that the odd nucleon is coupled to a triaxial core, leading to several coexisting rotational bands at low energy, which have near-degenerate band heads $J^\pi = 1/2^+$ (ground state), $J^\pi = 3/2^+$ (39.6 keV), $J^\pi = 11/2^-$ (236.1 keV), $J^\pi = 5/2^+$ (321.7 keV), $J^\pi = 7/2^+$ (665.4 keV), and a very similar spacing of the excited rotational levels~\cite{Timar14a}. Based on spectroscopic strength in $(^{3}\mathrm{He},x\mathrm{n})$ and $(\alpha,\mathrm{n})$ reactions when populating these levels, the band heads are usually associated with the spherical neutron $s_{1/2}$, $d_{3/2}$, $h_{11/2}$, $d_{5/2}$ and $g_{9/2}$ shells, respectively~\cite{Lonnroth83a,Timar14a}. A similar band structure is also found in adjacent odd-mass Xe isotopes, with the relative position of the band heads evolving with mass number~\cite{Banik20a}. In this context, the spectrum is usually explained either in terms of the Particle-Rotor Model (PRM) \cite{meyertervehn1975a} or the Interacting Boson-Fermion Model (IBFM) \cite{Cunningham82a,Cunningham82b,abu-musleh14a}. For both of these phenomenological models, parameters characterising the single-particle Hamiltonian of the spherical $j$-shells accessible to the odd nucleon, the collective Hamiltonian of the core, and the Hamiltonian describing the coupling of the core and the odd nucleon have to be determined in a nucleus-dependent manner. Traditionally, they all are fitted to the available data, but some of the parameters can also be deduced from other models. In some recent IBFM calculations for odd Xe isotopes, the parameters of the core and single-particle Hamiltonian are deduced from either a relativistic mean-field (RMF) model \cite{Nomura17a} or a non-relativistic HFB model using the Gogny force \cite{Nomura17b}, and only those of the Hamiltonian that couples the odd nucleon to the core are adjusted to data.

A common feature of the PRM and IBFM models is that the same core is used irrespective of the quantum numbers of the odd nucleon. This has to be contrasted with our MR-EDF approach that treats all nucleons on an equal footing without distinguish between valence nucleon and core.

\subsubsection{Generation of the reference states}
\label{sec:set129}

\begin{figure}[t!]
    \centering
    \includegraphics[width=0.70\linewidth]{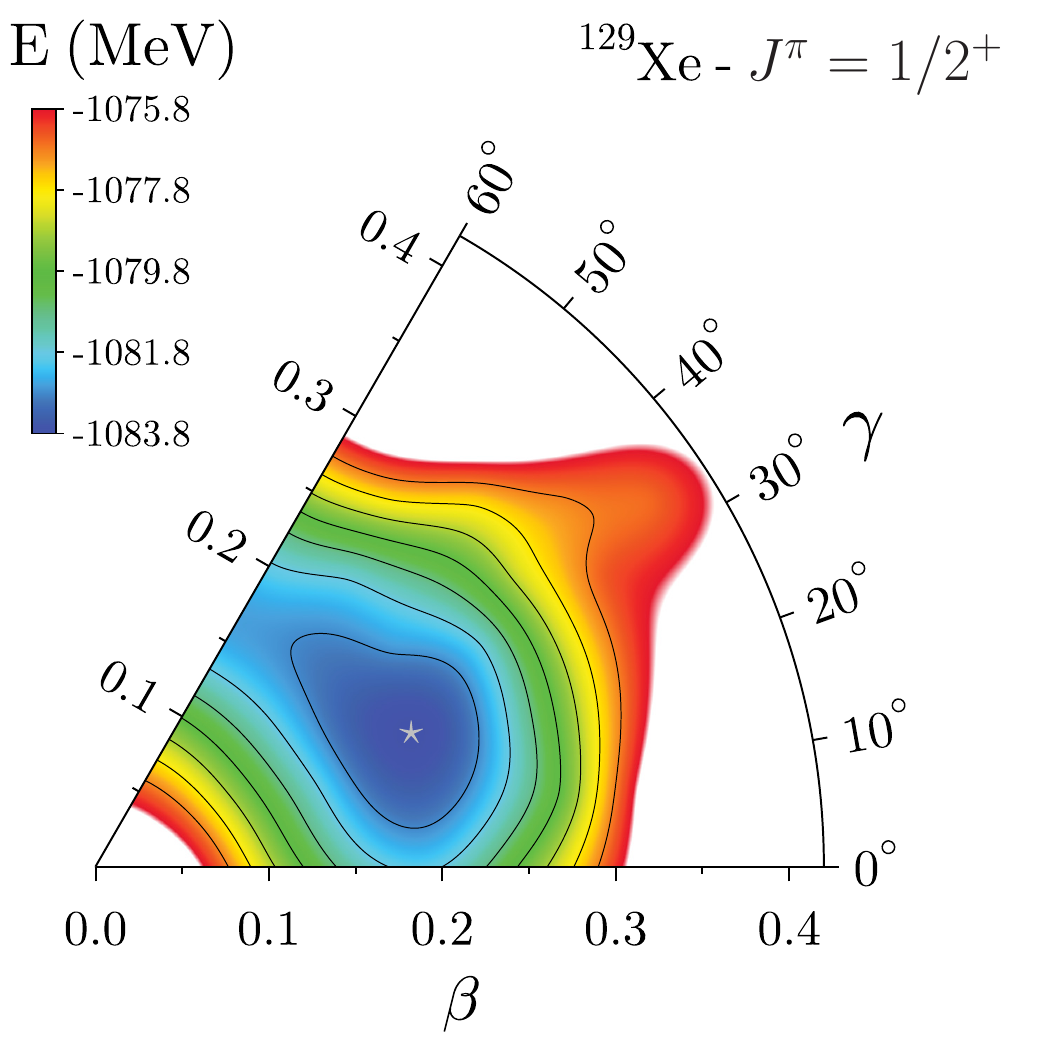} 
    \includegraphics[width=0.70\linewidth]{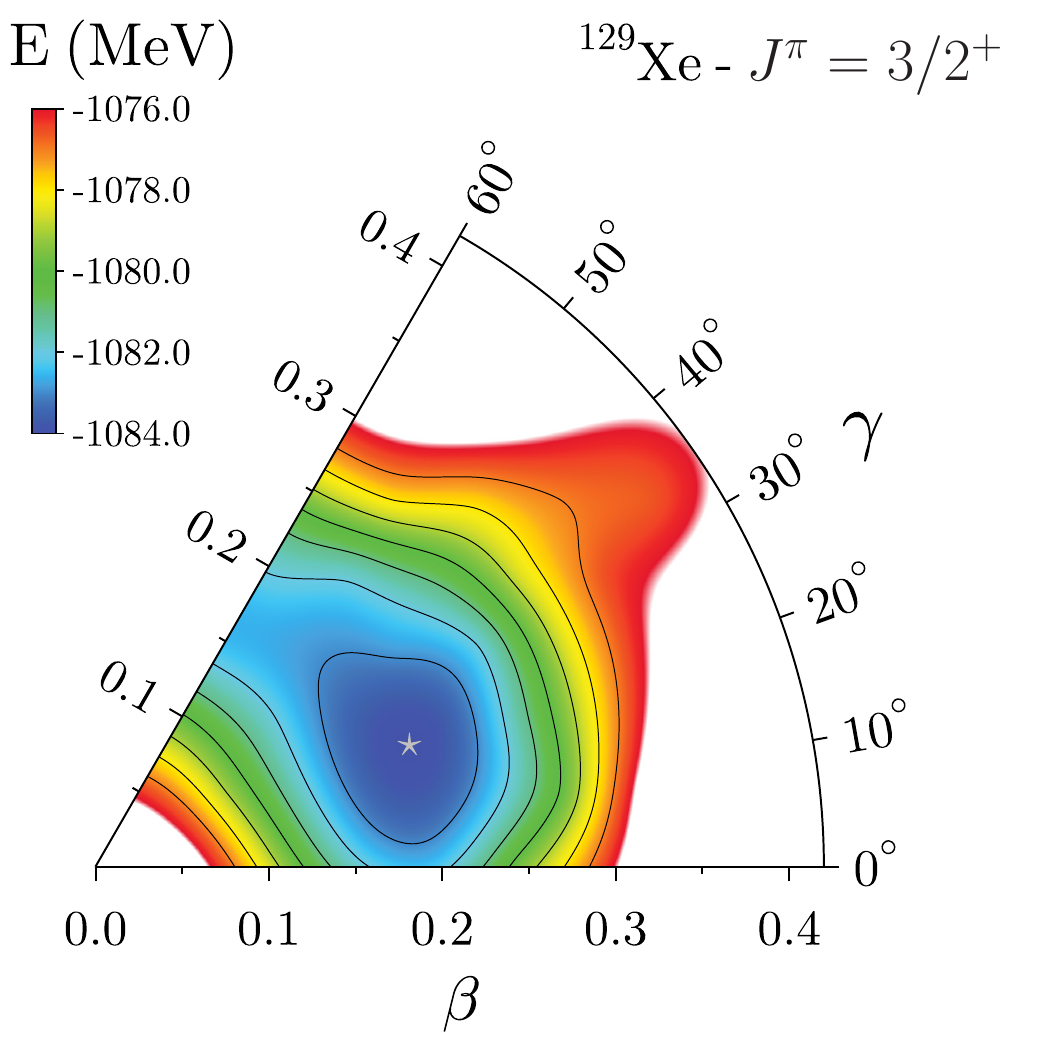}
    \caption{Same as Fig.~\ref{fig:xe128_ener_ampnr} but for $^{129}$Xe and the lowest $J^\pi = 1/2^+$ (top panel) or the lowest $J^\pi = 3/2^+$ (bottom panel).
    The minimum for $J^\pi =1/2^+$ ($J^\pi =3/2^+$), indicated by a silver star, is located at a deformation of $\beta=0.20$ and $\gamma=23^\circ$ ($\beta=0.19$ and $\gamma=21^\circ$).}
    \label{fig:xe129_ener_ampnr_plus}
\end{figure}

\begin{figure}[t!]
    \centering
    \includegraphics[width=0.70\linewidth]{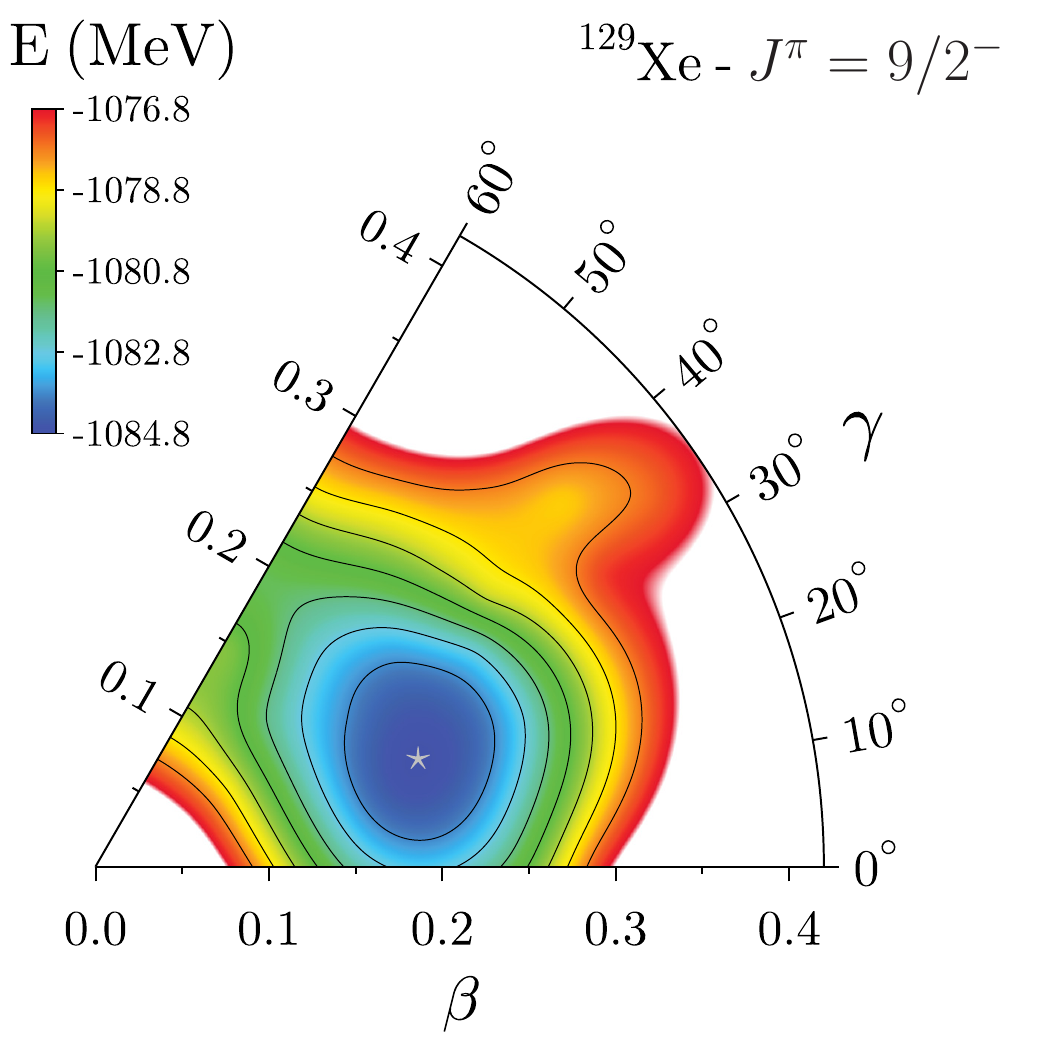} 
    \includegraphics[width=0.70\linewidth]{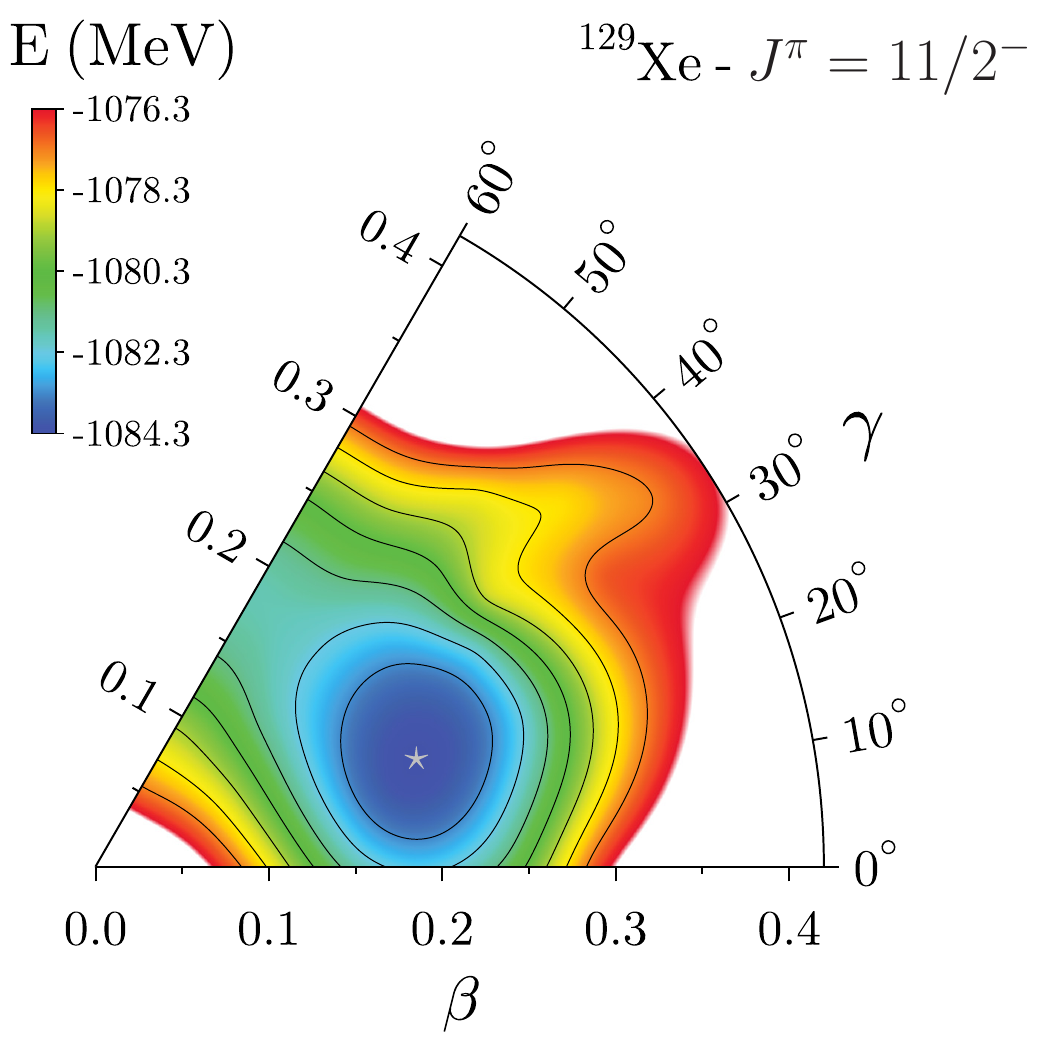} 
    \caption{Same as Fig.~\ref{fig:xe128_ener_ampnr} but for $^{129}$Xe and the lowest $J^\pi = 9/2^-$ (top panel) or the lowest $J^\pi = 11/2^-$ (bottom panel).
    The minimum for $J^\pi =9/2^-$ ($J^\pi =11/2^-$), indicated by a silver star, is located at a deformation of $\beta=0.20$ and $\gamma=18^\circ$  ($\beta=0.20$ and $\gamma=18^\circ$).}
    \label{fig:xe129_ener_ampnr_minus}
\end{figure}

In our MR framework, the treatment of odd-even nuclei is much more involved than the one of even-even nuclei. 
At each deformation, there exist several possibilities to block a one-quasi-particle state,
and it is not possible to guess with absolute certainty which configuration will be favored energetically at the end of the self-consistent procedure or at the beyond-mean-field level.
Therefore, in our calculations we adopt the following strategy:\footnote{We note that a different strategy was used to produced the results published in Ref.~\cite{Bally22a}. In those calculations we targeted specifically the lowest $1/2^+$ MR-EDF state and included only quasi-particle vacua that contain such components in their decomposition. Still, the two sets contain many common reference states and the final results obtained using one or the other are consistent.}
\begin{enumerate}
 \item For each deformation in our mesh, we converge an even vacuum constrained to have the correct odd number of particles on average as proposed in Ref.~\cite{Duguet01a}.
 \item For each such vacuum, we select the four one-quasi-particle states that have the lowest quasi-particle energy and block them self-consistently.
 \item We project all (converged) Bogoliubov states thus created and sort them according to the energy of their lowest projected state, without consideration for its angular momentum.
 \item For each parity, we select for the configuration mixing all the Bogoliubov states that give a projected energy lower than a given threshold above the lowest projected state of given parity.
\end{enumerate}
In the calculations of $^{129}$Xe described in what follows we use a threshold of 4~MeV for both positive and negative parities. This value is slightly lower than the one used when studying $^{128}$Xe, and is mainly motivated by the much larger number of reference states found at low excitation energies and the associated larger computational cost it implies. With this threshold, the configuration mixing includes 27 positive parity states and 35 negative parity states, which is enough to obtain a satisfying convergence of the various observables.

\subsubsection{Energy surfaces}
\label{sec:surface129}

Before analyzing the final configuration mixing, we plot in Fig.~\ref{fig:xe129_ener_pnr} the PNR total energy surface for both positive and negative parity states, selecting at each deformation the lowest particle-number projected state of a given parity. Both energy surfaces display a triaxial minimum and a topography similar to the PNR energy surface of $^{128}$Xe. But we note that in the negative parity case, the minimum is much closer to axiality.

In Figs.~\ref{fig:xe129_ener_ampnr_plus} and \ref{fig:xe129_ener_ampnr_minus}, we show the energy surfaces after subsequent projection of the reference states on angular momentum for $J^\pi = 1/2^+$, $3/2^+$, $9/2^-$, and $11/2^-$. We remark that we selected the lowest state with a given angular momentum and parity at each deformation and also that because we did not converge enough quasi-particle states with $\beta>0.3$, we display the energy surfaces only up to 8 MeV (contrary to 10 MeV in all other cases shown throughout the article) above the minimum. The similarity between the four surfaces is striking. The overall structure is equivalent in all four cases and the minima have very similar triaxial deformations. Indeed, the minima of negative parity states are pushed towards larger value of $\gamma$ compared to the simple PNR case. As the negative parity states gain more energy through angular momentum restoration then the positive parity states (approximately 4 MeV for the $9/2^-$ surface compared to 3.6 MeV for the $3/2^+$ surface), the slightly wrong relative position (approximately 200 keV) of the PNR minima becomes very much amplified. As a consequence, we obtain for the angular-momentum projected minima the relative order: $9/2^-$, $11/2^-$, $3/2^+$, $1/2^+$, which is almost the complete opposite of what is observed experimentally. 

\subsubsection{Low-energy spectroscopy}
\label{sec:spectrum129}

\begin{figure}[t!]
    \centering
    \includegraphics[width=.80\linewidth]{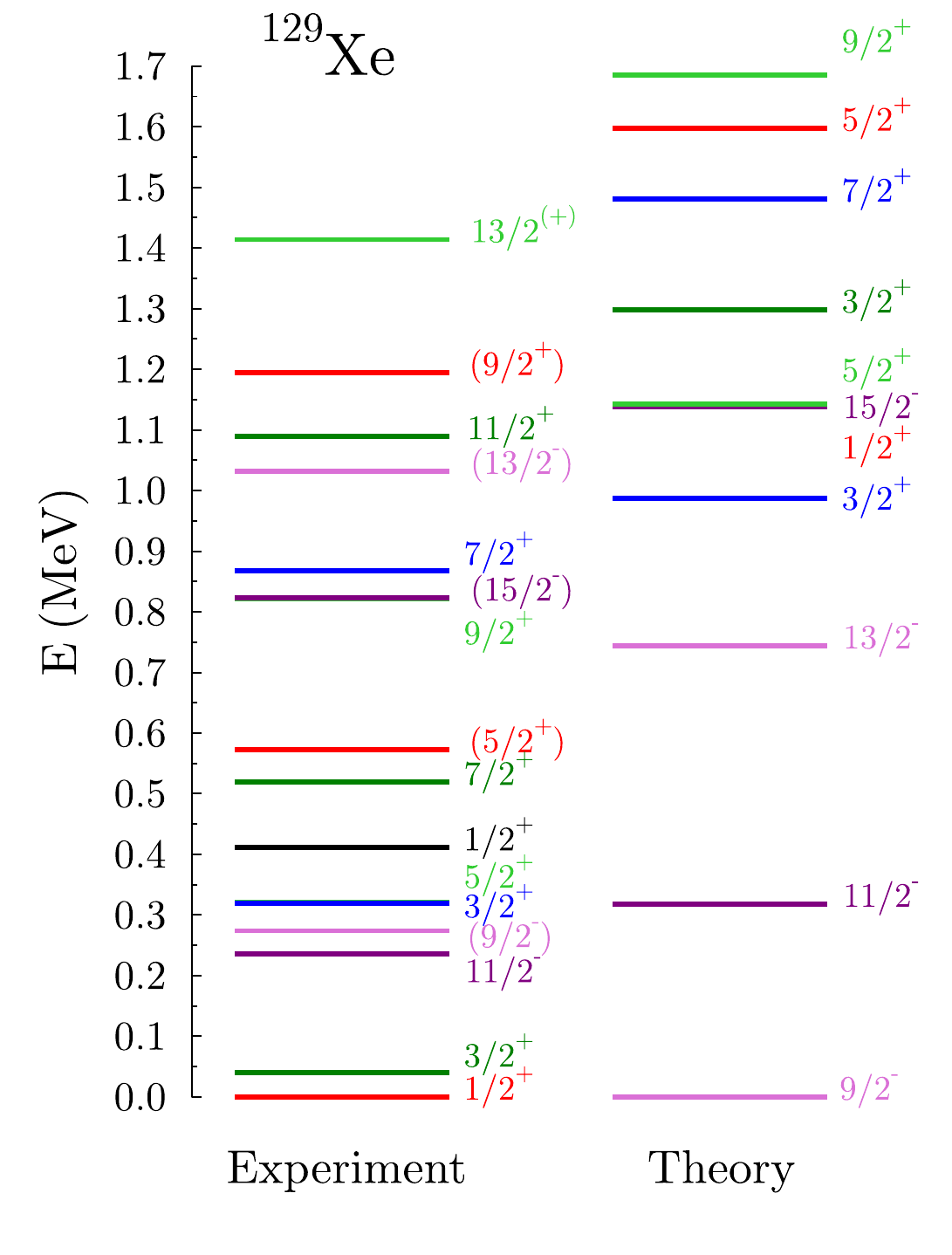}
    \caption{Low-energy spectrum for $^{129}$Xe. We mention that the experimental (theoretical) $3/2_2^+$ and $5/2_1^+$ as well as $9/2_1^+$ and $15/2_1^-$ ($1/2_1^+$, $15/2_1^-$ and $5/2_1^+$) states appear as degenerate with the energy scale used in the figure. Experimental data are taken from \cite{AIEAa,Timar14a}, where levels that are grouped into bands based on the same quasi-particle are drawn in the same color, with levels in the favoured (disfavoured) bands drawn in dark (light) color tones.
    }
    \label{fig:xe129_spec}
\end{figure}

The configuration mixing confirms this order as can be seen in the low-energy spectrum displayed in Fig.~\ref{fig:xe129_spec}. If all the states\footnote{We stress here that we only display a few out of the low-lying states known experimentally as well as their theoretical counterparts.} that appear experimentally up to $\simeq$ 500 keV are present in our calculations, their relative order is evidently not reproduced at all. First, the calculated ground state is a $9/2^-_1$ state instead of a $1/2^+_1$. Second, the order among the pairs ($9/2^-_1$, $11/2^-_1$) and ($1/2^+_1$, $3/2^+_1$) are inverted in our results compared to experimental data. But the most striking discrepancy is the large gap in energy between the negative and positive parity states.

As mentioned above, the low-lying part of the excitation spectrum is usually interpreted in terms of coexisting rotational bands whose band heads are near-degenerate. The precise reproduction of such a spectrum is beyond of what can be expected from present nuclear EDF methods~\cite{Bonneau07a}, in particular when also considering that the theoretical single-particle states associated with the band heads can mix. The authors of Refs.~\cite{Nomura17a,Nomura17b} point out that the distances between single-particle energies obtained with the relativistic DD-PC1 and Gogny D1M parameterizations that enter their EDF-based IBFM calculations significantly disagree with each other and the values used in the earlier purely phenomenological calculations on the order of several hundreds of keV, sometimes more than 1~MeV. This required them to introduce a $j$-dependent monopole shift in their Hamiltonian in order to reproduce the excitation spectra.
The deviation of our results is, however, larger than what is typically found and therefore merits some further analysis.

The differences go in fact further than just an inversion of levels. To ease the comparison of the spectra, levels attributed to different rotational bands are drawn in different colours in Fig.~\ref{fig:xe129_spec}, where in addition the favoured bands are represented with dark tones, and the disfavoured bands based on the same band heads with light tones. For the experimental data we follow Ref.~\cite{Timar14a}, and the grouping of calculated levels into bands is based on relative strength of $E2$ transitions and spectroscopic quadrupole moments, and their similarity with experimental data. 

\begin{table}
\centering
\begin{tabular}{ccc}
\hline\noalign{\smallskip}
Quantity & Experiment & Theory \\
\noalign{\smallskip}\hline\noalign{\smallskip}
$E(1/2^+_1)$ & -1087.649 & -1084.720 \\[0.08cm]
$r_\text{rms}(1/2^+_1)$ & 4.7775(50) & 4.736 \\[0.08cm]
$\mu(1/2^+_1)$ & -0.78 & -0.2 \\[0.08cm]
$\mu(3/2^+_1)$ & +0.58(8) & +1.2 \\[0.08cm]
$\mu(3/2^+_2)$ &           & -0.1 \\[0.08cm]
$\mu(11/2^-_1)$ & -0.89 &  -1.1 \\[0.08cm]
$Q_s(3/2^+_1)$ & -0.39(1) & +0.4 \\[0.08cm]
$Q_s(3/2^+_2)$ &           & -0.4 \\[0.08cm]
$Q_s(11/2^-_1)$ & +0.63(2) & +0.8 \\
\noalign{\smallskip}\hline
\end{tabular}
\caption{Same as Table \ref{tab:values128} but for $^{129}$Xe. Experimental data are taken from \cite{Angeli13a,Stone16a,Stone19a,Stone20a,Huang21a,Wang21a}. We note that the magnetic moments of the $1/2^+_1$ and $11/2^-_1$ states are known experimentally up to an extremely good accuracy and were truncated here at the second digit.}
\label{tab:values129}
\end{table}

In Table \ref{tab:values129}, we report spectroscopic quantities for some of the low-lying states. First, we notice that even though we obtain the wrong order of low-lying levels, the total energy of the $1/2^+_1$ state is only 3 MeV off the experimental value. Similarly, the root-mean-square charge radius of this state is reproduced with an accuracy similar to the one of the $0_1^+$ state of the $^{128}$Xe nucleus. On the other hand, the electromagnetic moments are not all as well described. The calculation reproduces the spectroscopic quadrupole moments $Q_s$ of the lowest $3/2^+$ state, provided that we interpret the order of the two lowest $3/2^+$ states in the calculated spectrum as incorrectly reversed compared to experimental data. The $Q_s$ of the $11/2^-_1$ state is also reasonably close to the data, but slightly overestimated, which for a triaxial states does not necessarily indicate too large overall deformation, but could also point to an incorrect $K$ mixing. The magnetic moment of the $11/2^-_1$ state is also slightly overestimated by about 20$\%$. The calculated values of the low-spin positive parity states, however, differ enormously from the experimental ones, in particular considering when assuming that the second calculated $3/2^+$ state should be compared with the first experimental $3/2^+$ state and vice versa.

Nuclear EDF methods usually perform quite well for electric quadrupole moments, such that large differences with experimental data beyond about ten percent can usually be attributed to some deficiency of the description of a given nuclear configuration. We have to recall, however, that it is much less clear how to interpret large differences between calculated and experimental magnetic moments. Indeed, there are several reasons why magnetic moments are notoriously difficult to describe theoretically, which can easily result in large differences of several tens of percent, even a wrong sign when the absolute value is small. This overall poor performance explains why in the past magnetic moments have been rarely analysed in the context of pure EDF methods. The main reason for these difficulties is that magnetic moments sensitively probe the distribution of current and spin in nuclei. Unlike their contributions to the electric charge density that mostly add up, the individual contributions of nucleons to current and spin tend to cancel each other. What remains is in general a combination of the contributions from the single-particle states of the blocked nucleons, from the collective motion of the other nucleons, and from polarisation effects induced by blocking and collective motion. In particular the latter is difficult to control in nuclear models as this effect is dominated by the so-called time-odd terms that contribute only very little to energetic observables and therefore are not well constrained in the parameter adjustment. Recent MR-EDF studies limited to angular-momentum projected axial configurations~\cite{Sassarini21a,Vernon22a} propose to improve the description of magnetic moments through a fine-tuning of time-odd terms, but it remains unclear if their contribution is the major source for the discrepancies between experiment and theory.

We also recall that in approaches that rely on valence spaces, including the PRM and IBFM mentioned above, these difficulties are at least partially masked by the use of effective $g$-factors that are adjusted to reproduce global trends of data in the $M1$ operator 
\begin{equation}
\label{eq:M1:1body}
\hat{M}_{1\mu}
= \sum_{i=1}^{A} 
  \Big( g^{(\ell)}_{i} \, \hat{\ell}_{i,\mu} 
       +g^{(s)}_{i} \, \hat{s}_{i,\mu} 
  \Big) \, \mu_N \, ,
\end{equation}
where $\hat{\ell}_{i,\mu}$ and $\hat{s}_{i,\mu} $ are the $\mu \in \{ x, y, z\}$ components of the orbital angular momentum and spin operators of the $i$-th nucleon, respectively, and $\mu_N$ the nuclear magneton. In EDF approaches that construct the reference states in the full space of occupied particles as the one employed here, there is no justification to modify the $M1$ operator for reasons of limited valence space in the actual EDF calculation and therefore we use the bare $g$-factors of protons and neutrons in Eq.~\eqref{eq:M1:1body} instead. There is, however, a related issue with using an EDF in the first place: In an effective manner, the EDF resums in-medium correlations and some finite-size correlations, and the contributions from configurations that are implicitly integrated out in the energy should also modify all other operators. In addition, the part of the nuclear Hamiltonian that changes the nucleon's isospin (and thereby the nucleon's electromagnetic properties) generates two-body (and higher-order) currents that the magnetic fields also couple to \cite{Ericson88a}. In an \textit{ab initio} approach, one can in principle construct a consistent $M1$ operator that takes these two effects into account. In the absence of a constructive scheme that establishes a nuclear EDF from first principles it is not clear how to introduce them other than phenomenologically, which, however, with few exceptions such as Ref.~\cite{Borzov08a,Co15a} is also not systematically done. These considerations also might be at the origin of the discrepancy between calculated and experimental $M1$ transition strength between the levels of $^{128}$Xe mentioned above.

\begin{figure}[t!]
    \centering
    \includegraphics[width=.90\linewidth]{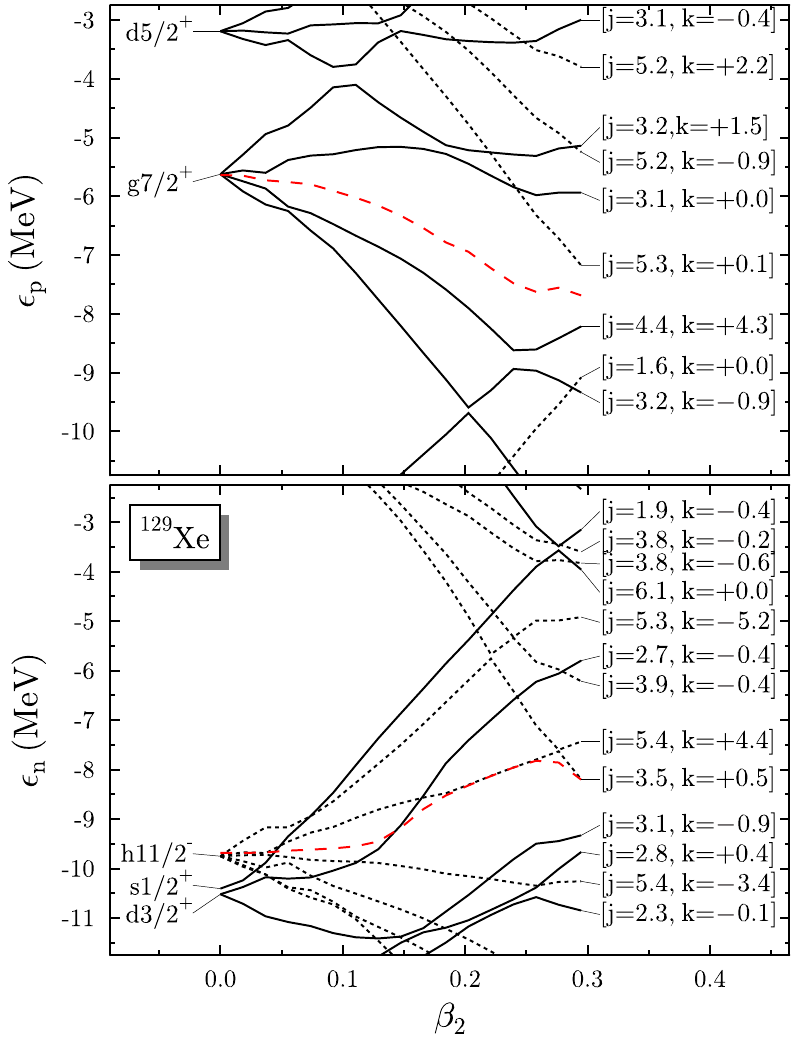}
    \caption{Nilsson diagram of the eigenvalues of the single-particle Hamiltonian for protons and neutrons obtained for false HFB vacua constrained to $Z=54$, $N=75$ and $\gamma = 19^\circ$ as a function of $\beta$ as given by SLyMR1. Solid (dotted) black lines are used to draw levels of positive (negative) parity, whereas the dashed red lines indicate the Fermi energies of protons and neutrons, respectively. At $\beta = 0$, labels indicate quantum numbers of spherical $j$ shells, whereas labels at the largest $\beta$ indicate the average values of $j$ obtained from $\langle \hat{\jmath}^2 \rangle = j(j+1)$ and of $\langle \hat{\jmath}_z \rangle = k$ (see text for the interpretation of the latter's sign).
    }
    \label{fig:xe129_nilsson}
\end{figure}

Figure~\ref{fig:xe129_nilsson} shows the Nilsson diagrams of eigenvalues of the single-particle Hamiltwonian obtained with SLyMR1 for false vacua of $^{129}$Xe along a cut at $\gamma = 19^\circ$ through the $\beta$-$\gamma$ plane that passes near the deformations of dominating contributions to the low-lying collective states. Labels to the left indicate quantum numbers of spherical $j$ shells, labels to the right indicate the average values of total angular momentum $j$ obtained as $\langle \hat{\jmath}^2 \rangle = j(j+1)$ and of the $z$ component of total angular momentum $ \langle \hat{\jmath}_z \rangle = k$. The indicated expectation value of $\hat{\jmath}_z$ is calculated form the state with positive signature out of the Kramers-degenerate pair. For axial states that are aligned with the $z$ axis, $k$ takes the values $+1/2$, $-3/2$, $+5/2$, $-7/2$, such that for these non-axial states the sign of $k$ gives an indication of the dominating components. The overall appearance of the spectra is similar to what has been reported for nearby Xe isotopes with a Woods-Saxon potential \cite{Huang16a}, the DD-PC1 point-coupling RMF Lagrangian \cite{Nomura17a} or the D1M parameterisation of the Gogny force \cite{Nomura17b}, with small differences between all of them concerning the placement of the spherical shells.

The dominant contributions to the collective wave function (see below) are not exactly at the $\gamma$ value used to prepare Fig.~\ref{fig:xe129_nilsson}, and we also recall that HFB and self-consistent blocking mix configurations, but it can be easily seen that the energetically lowest blocked quasi-particle vacuum is based on the single-particle level from the $h_{11/2}$ shell whose mean value of $\langle \hat{\jmath}_z \rangle$ remains close to $9/2$, and that the lowest states of positive parity will be dominated by the upsloping single-particle level just above the Fermi energy labeled by $j = 2.7$ and $k = -0.4$.
All neutron single-particle levels of positive parity that are close to the Fermi surface at deformations of $\beta \approx 0.2$, which  are close to the minima of the energy surfaces shown in Fig.~\ref{fig:xe129_ener_ampnr_plus}, have similar mean values of $j$ between 2 and 3 and similar absolute mean values of $k$ around 0.5, in spite of their originating from the spherical $s_{1/2}$, $d_{3/2}$ and (for the ones upsloping from below into the part of the spectrum that is shown) $d_{5/2}$ shells, indicating that they are all very mixed by the quadrupole deformation. The smallness and negative sign of the average $k$ for the single-particle state of positive parity that is closest to the Fermi energy indicates that it is dominated by a $k=3/2$ component. Blocking this level generates quasi-particle vacua that 
after projection and configuration mixing dominate the $3/2^+_1$ level in Fig.~\ref{fig:xe129_spec} and explains why the latter is the lowest positive-parity state in the calculated spectrum, at variance with experiment. The respective components projected from the same quasi-particle vacua dominate also the $1/2^+_1$ and $5/2^+_1$ levels. Similarly, the lowest $9/2^-$ and $11/2^-$ MR states in the calculated spectrum of Fig.~\ref{fig:xe129_spec} are also dominated by components that at a given deformation are projected out from the same blocked quasi-particle configuration, in this case the one obtained blocking the negative-parity neutron $k \approx 9/2$ level at the Fermi surface. Again, projection and mixing gives the wrong order of the $9/2^-_1$ and $11/2^-_1$ levels. 

\begin{figure}[t!]
    \centering
    \includegraphics[width=0.70\linewidth]{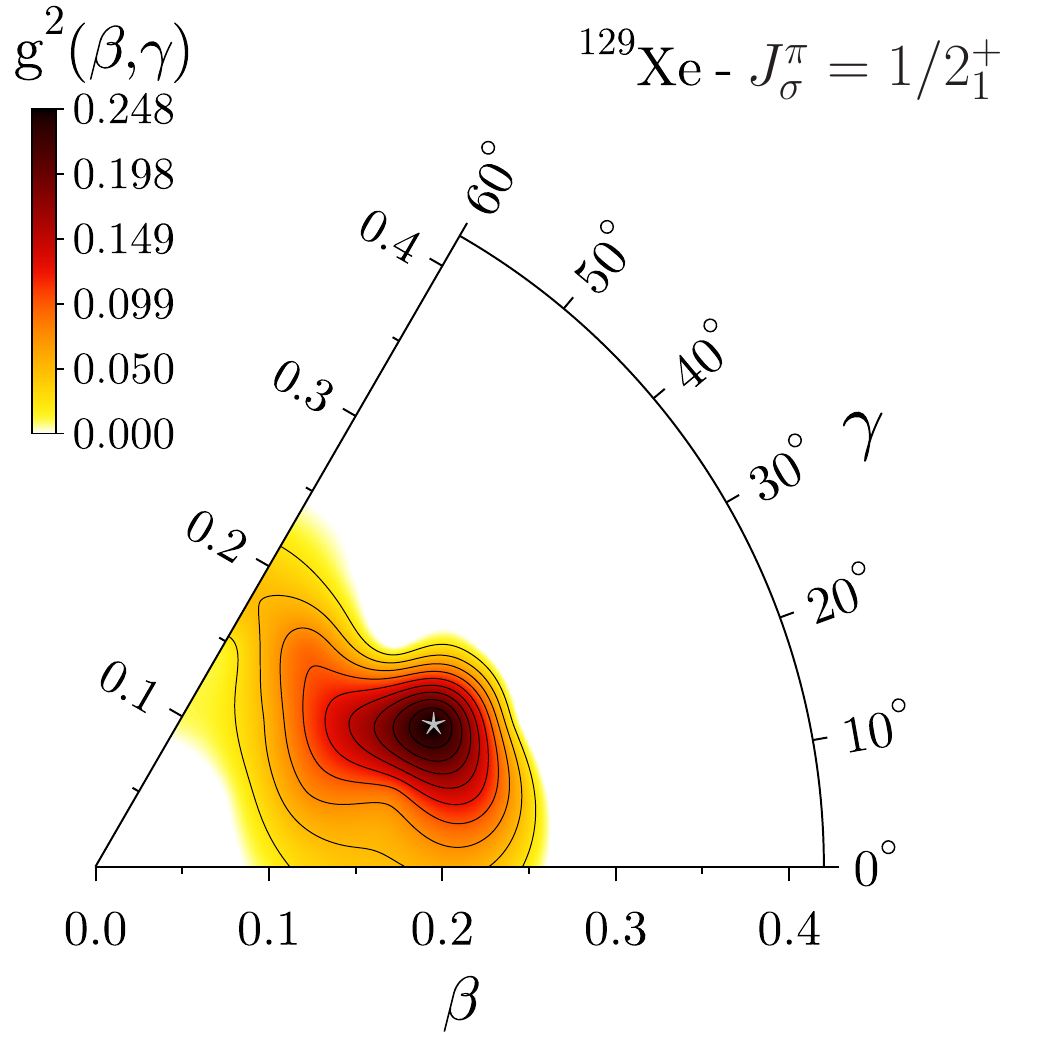}
    \includegraphics[width=0.70\linewidth]{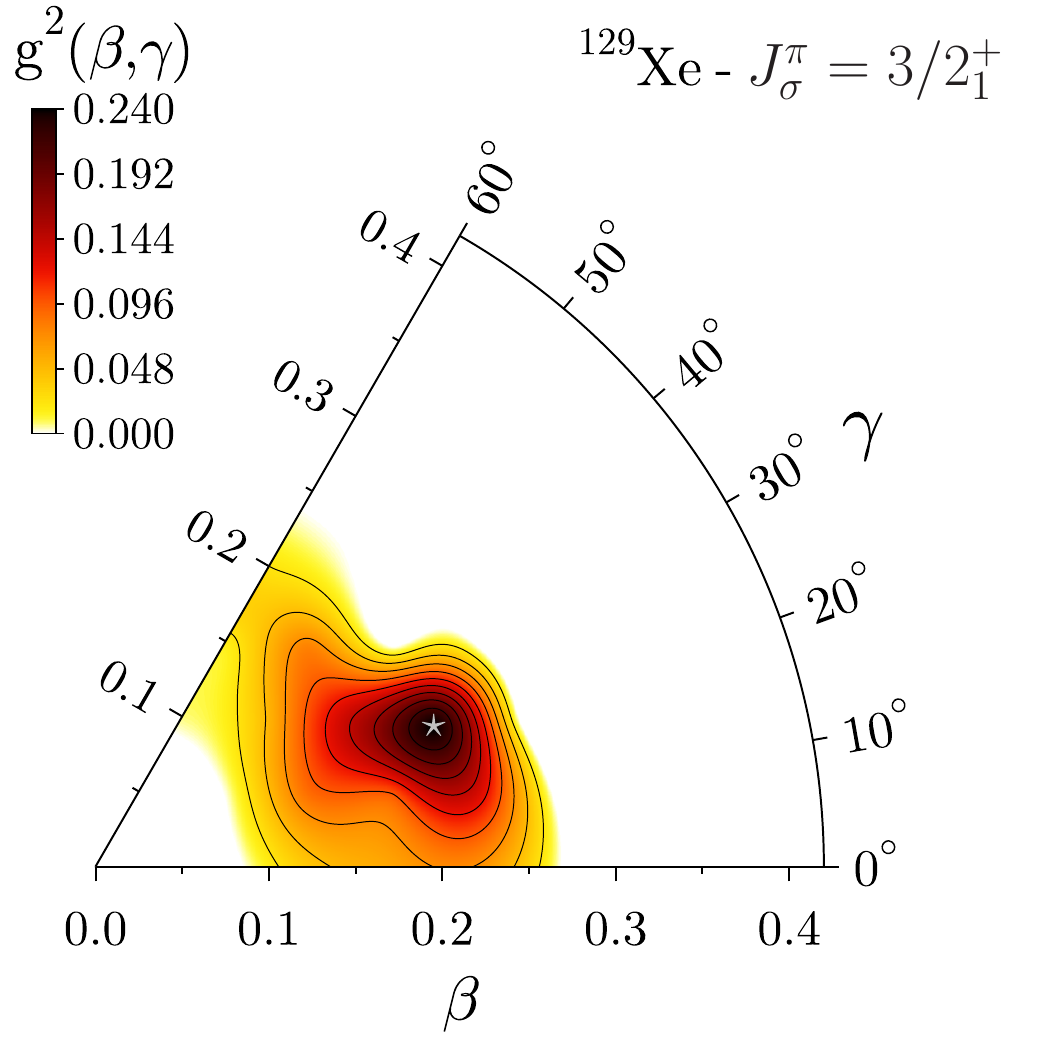}
    \caption{Same as Fig.~\ref{fig:xe128_coll} but for $^{129}$Xe and  $J^\pi_\sigma=1/2^+_1$ (top panel) or $3/2^+_1$ (bottom panel).
    The maximum for $J^\pi_\sigma=1/2^+_1$ ($3/2^+_1$), indicated by a silver star, is located at a deformation of $\beta=0.21$ and $\gamma=23^\circ$ ($\beta=0.21$ and $\gamma=23^\circ$).}
    \label{fig:xe129_plus_coll}
\end{figure}

\begin{figure}[t!]
    \centering
    \includegraphics[width=0.70\linewidth]{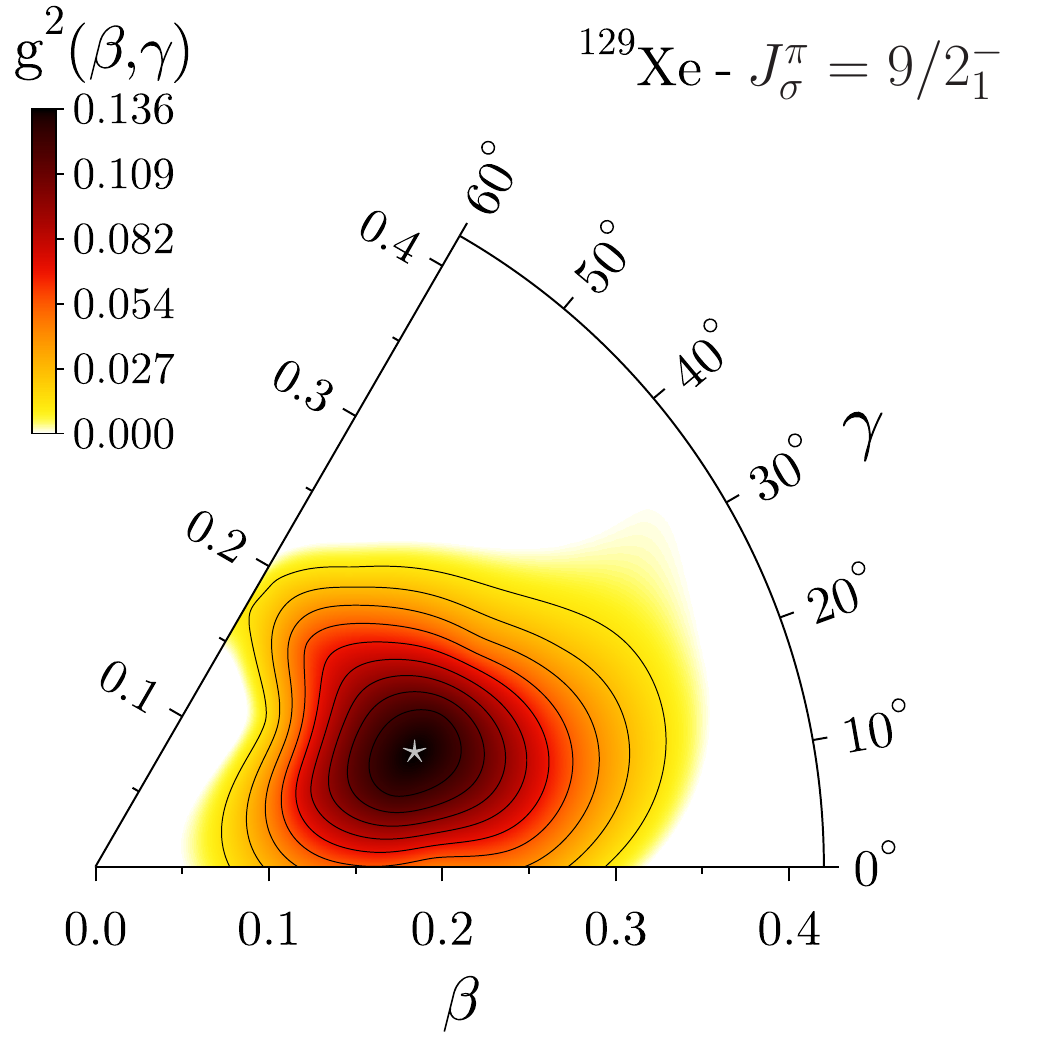}
    \includegraphics[width=0.70\linewidth]{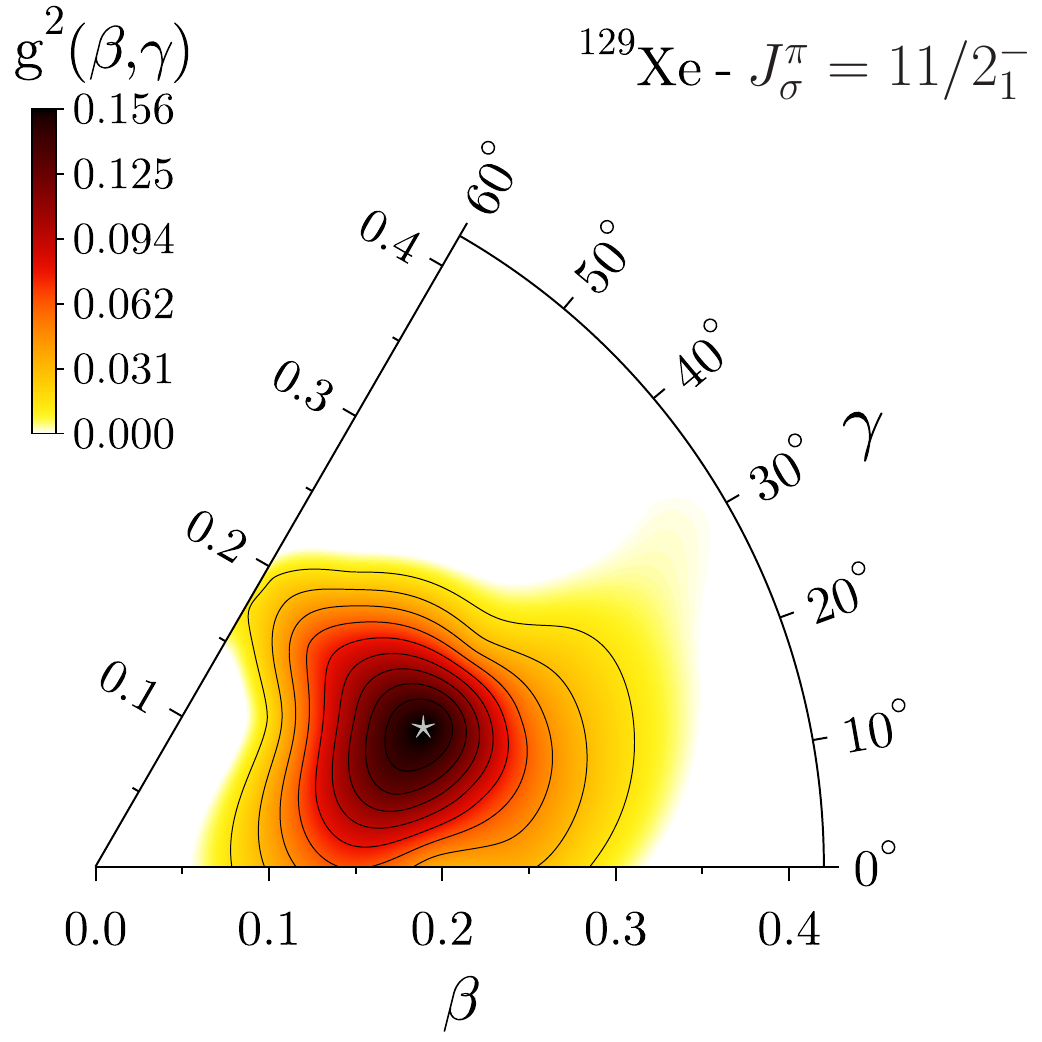}
    \caption{Same as Fig.~\ref{fig:xe128_coll} but for $^{129}$Xe and  $J^\pi_\sigma=9/2^-_1$ (top panel) or $11/2^-_1$ (bottom panel).
    The maximum for $J^\pi_\sigma=9/2^-_1$ ($11/2^-_1$), indicated by a silver star, is located at a deformation of $\beta=0.20$ and $\gamma=20^\circ$ ($\beta=0.21$ and $\gamma=23^\circ$).}
    \label{fig:xe129_minus_coll}
\end{figure}

Supposing that all bands of $^{129}$Xe have similar triaxial deformation and assuming that the relative position and average value $\langle \hat{\jmath}_z \rangle$ of single-particle levels in the Nilsson diagram at $\beta \approx 0.2$ and $\gamma \approx 20^\circ$ is the determining factor for the relative positions of the band heads in the MR calculation of this nucleus, then the reason for the incorrect description of the low-lying spectrum of this nucleus can be attributed to the wrong relative position of the neutron shells at sphericity: the $h_{11/2}$ level has to be lowered so much that all three positive-parity levels from the $s_{1/2}$ and $d_{3/2}$ are above the $k \approx 9/2$ level from the $h_{11/2}$ shell at the deformations that dominate the collective state, and the relative position of the $s_{1/2}$ and $d_{3/2}$ shells has to be changed in such a way that the Kramers-degenerate deformed level from the $s_{1/2}$ shell remains comparatively unmixed and is the lowest of the three positive-parity levels at the deformation of the band heads.

\subsubsection{Average deformation}
\label{sec:avgdef129}

Lastly, we analyze the cwf for the low-lying calculated $9/2^-_1$, $11/2^-_1$, $1/2^+_1$, $3/2^+_1$ states displayed in Figs.~\ref{fig:xe129_plus_coll} and \ref{fig:xe129_minus_coll}. The squared cwf (scwf) of a same parity are almost identical while we see differences between the opposite parities. In particular, the scwf of negative parity states are much wider and extends towards larger value of $\beta$ at small values of $\gamma$. Still, in all cases the scwf is centered around a sharply peaked triaxial maximum with a deformation of roughly $\beta \approx20 $ and $\gamma \approx 20\textrm{--}\,25^\circ$. Interestingly, the scwf of the $3/2^+_2$ states (not shown here) is also extremely similar to the one of $3/2^+_1$ but when analyzing their decomposition in terms of $K$-components along the lines of Ref.~\cite{Bally21a} we find that they have almost opposite behavior. Indeed, the most important contribution in the decomposition of the $3/2^+_1$ comes from the $K=3/2$ component, whereas for the $3/2^+_2$ state, it is the $K=1/2$ component that dominates. This observation is consistent with the analysis of the single-particle states discussed above.

Also, this explains the opposite quadrupole moment mentioned above. Indeed, making the somewhat oversimplifying assumption that the contributions of the matrix elements of $\hat{E}_{22}$ are negligible and that the collective states have only one $K$ component,
the rigid-rotor model expression for the spectroscopic quadrupole moment $Q_s$ of such states \cite{RS80a} 
\begin{equation}
Q_s
= Q_0 \, \frac{3 K^2 - J(J+1)}{(J+1)(2J+3)}
\end{equation}
gives values of opposite sign for $J=3/2$ states with $K=1/2$ and $3/2$.

Using the scwf of the $1/2_1^+$ state, we are able to compute the average elongation $\beta_{c}(1/2^+_{1}) = 0.18$, with a standard deviation $\Delta \beta_c(1/2^+_{1}) = 0.03$, and the average angle $\gamma_{c}(1/2^+_{1}) = 25^\circ$, with a standard deviation $\Delta \gamma_c(1/2^+_{1}) = 16^\circ$. Reassuringly, these values are in good agreement with the ones of our previous calculations reported in Ref.~\cite{Bally22a} and that differed by the set of reference states used.

Given the poor quality of the calculated spectrum displayed in Fig.~\ref{fig:xe129_spec}, we did not compute the Kumar parameters for the $1/2_1^+$ state of $^{129}$Xe. In addition, the presence at low energy of many rotational bands associated with different single-particle configurations calls into question the possibility to connect  in a meaningful way the matrix elements of $E2$ transitions among the states to the moments of a collective quadrupole shape.


\subsection{$^{130}$Xe}
\label{sec:130}

Finally, we turn our attention towards $^{130}$Xe, which is of particular interest as it was the focus of a recent Coulomb excitation experiment \cite{Morrison20a}. Our methodology is identical to the one used to study the lighter even-mass isotope $^{128}$Xe. 


\subsubsection{Energy surfaces}
\label{sec:surface130}

\begin{figure}[t!]
    \centering
    \includegraphics[width=0.70\linewidth]{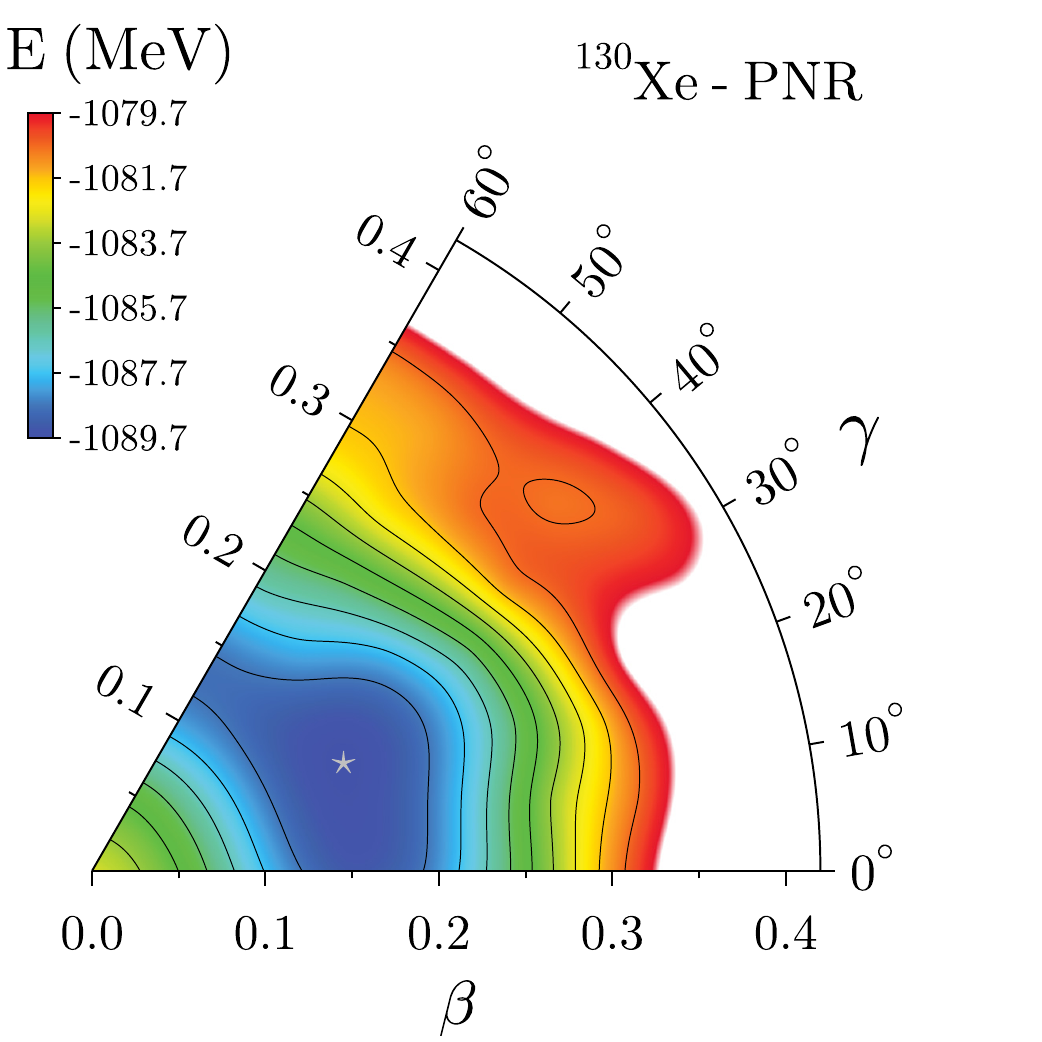} 
    \caption{Same as Fig.~\ref{fig:xe128_ener_pnr} but for $^{130}$Xe. 
    The minimum, indicated by a silver star, is located at a deformation of $\beta=0.16$ and $\gamma=23^\circ$.}
     \label{fig:xe130_ener_pnr}
\end{figure}

\begin{figure}[t!]
    \centering
    \includegraphics[width=0.70\linewidth]{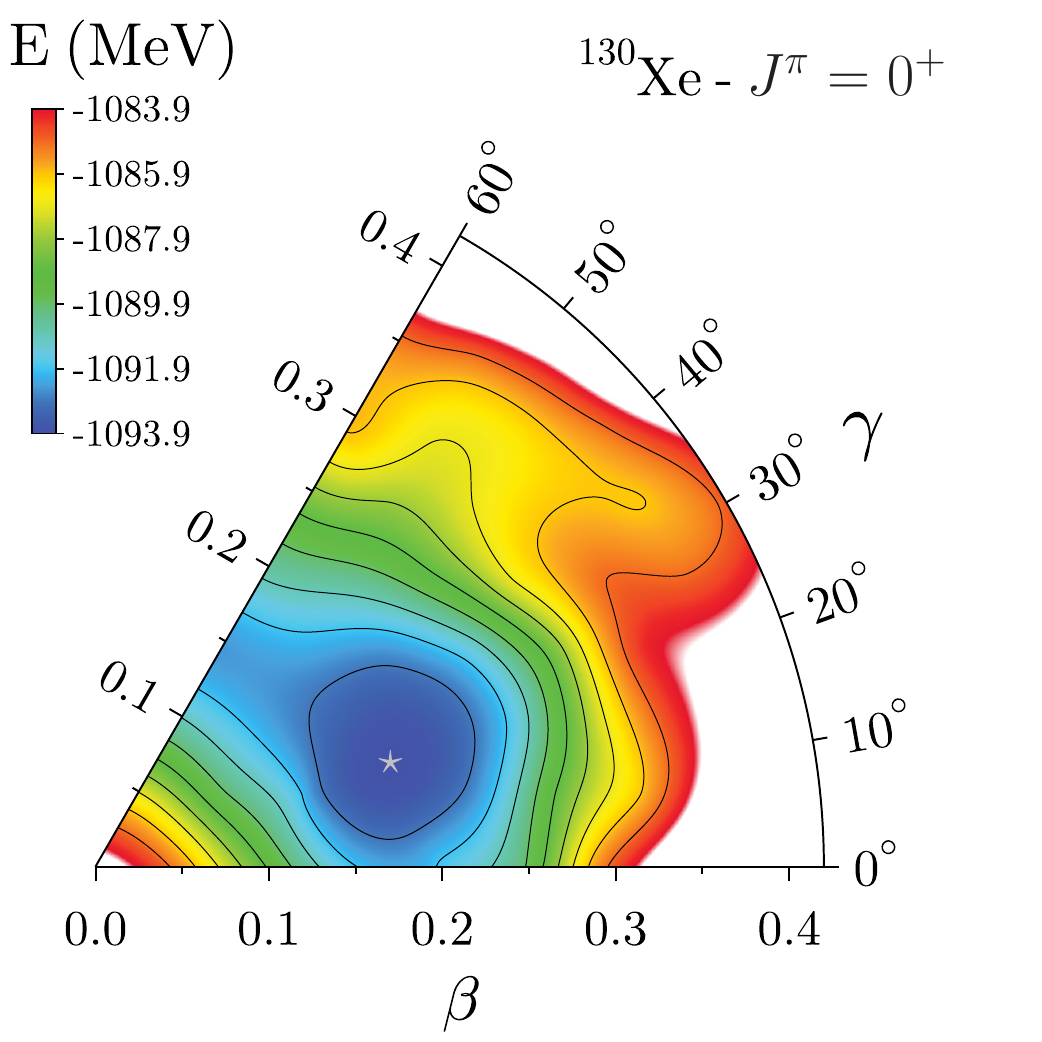} 
    \caption{Same as Fig.~\ref{fig:xe128_ener_ampnr} but for $^{130}$Xe.
    The minimum, indicated by a silver star, is located at a deformation of $\beta=0.18$ and $\gamma=19^\circ$.}
    \label{fig:xe130_ener_ampnr}
\end{figure}

In Figs.~\ref{fig:xe130_ener_pnr} and \ref{fig:xe130_ener_ampnr}, we display the PNR and $J^\pi = 0^+$ total energy surfaces, respectively. At the PNR level, the surface is more rigid (softer) against $\beta$- ($\gamma$-) deformation than in the case of $^{128}$Xe. In addition, the particle-number and angular-momentum projected minimum, which is lowered by approximately 4 MeV compared to a calculation where only particle numbers are restored, is located at the deformation of $\beta = 0.16$ and $\gamma = 23^\circ$ and, therefore, is slightly less deformed and slightly more triaxial. Once the rotational invariance is restored, the surfaces of the two nuclei become more similar, although the more rigid character against $\beta$-deformation remains in the case of $^{130}$Xe. 


\subsubsection{Low-energy spectroscopy}
\label{sec:spectrum130}

\begin{figure}[t!]
    \centering
    \includegraphics[width=.80\linewidth]{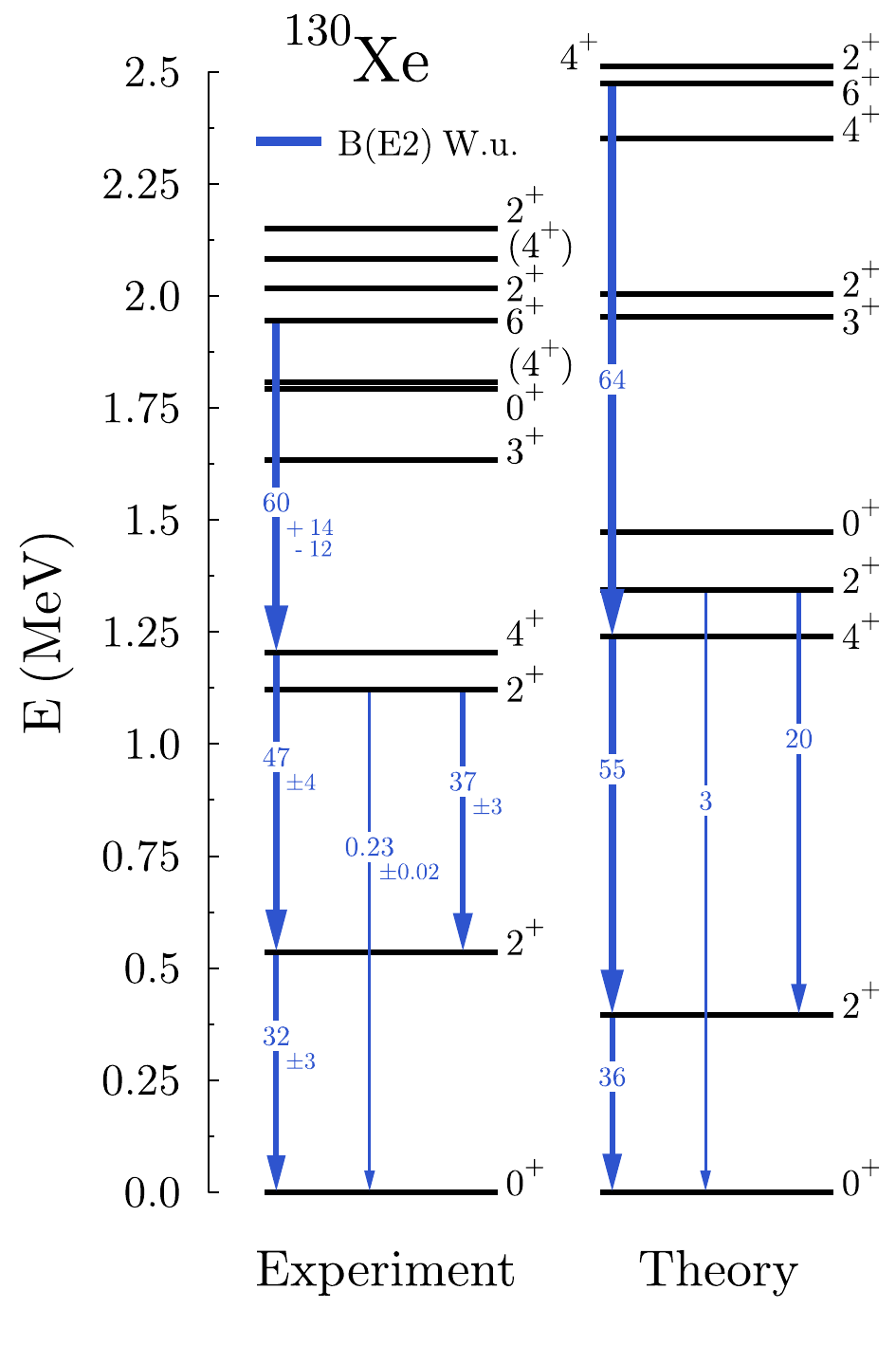}
    \caption{Same as Fig.~\ref{fig:xe128_spec} but for $^{130}$Xe. Experimental data are taken from \cite{Morrison20a,AIEAa,Singh01a}.}
    \label{fig:xe130_spec}
\end{figure}

Performing the configuration mixing, which in this case includes 22 states, we show in Fig.~\ref{fig:xe130_spec} the energy spectrum for $^{130}$Xe. We obtain a very good description of the lowest part of the spectrum, below 1.25 MeV, while the higher-lying states are somewhat too spread. As in the case of $^{128}$Xe, the $2^+_2$ and $4^+_1$ states are inverted in our calculations compared to the experimental observations. In contrast to $^{128}$Xe, the calculated $0^+_2$ state is too low in energy for $^{130}$Xe. Concerning the electromagnetic decays, the intensity of the known transitions are well reproduced. 
Overall, we achieve a reasonable (but improvable) description of the energy spectrum of $^{130}$Xe that is on a par with the quality of the shell-model calculations discussed in Ref.~\cite{Morrison20a}. We mention in passing that $^{130}$Xe was also studied within the IBM framework and the features of its low-energy spectrum were shown to be compatible with the picture of a near $O(6)$ nucleus \cite{Arima79a,Casten85a}, suggesting again a triaxially deformed nucleus.

In Table \ref{tab:values130}, we report spectroscopic quantities for some of the low-lying states. We obtain a fairly good agreement for the binding energy, the root-mean-square charge radius as well as for the electric quadrupole and magnetic dipole moments of the $2^+_1$ and $4^+_1$ states. The results are somewhat less close for the $2^+_2$ state, with in particular a larger overestimation of its spectroscopic quadrupole moments. Nevertheless, we obtain the correct sign in every single case.

\begin{table}
\centering
\begin{tabular}{ccc}
\noalign{\smallskip}\hline
Quantity & Experiment & Theory \\
\noalign{\smallskip}\hline\noalign{\smallskip}
$E(0^+_1)$ & -1096.905 & -1095.058 \\[0.08cm]
$r_\text{rms}(0^+_1)$ & 4.7818(49) & 4.741 \\[0.08cm]
$\mu(2^+_1)$ & +0.67(10)  & +0.6 \\[0.08cm]
$\mu(4^+_1)$ & +1.7(4) & +1.3 \\[0.08cm]
$\mu(2^+_2)$ & +0.9(2)  & +0.6 \\[0.08cm]
$Q_s(2^+_1)$ & -0.38(+17,-14) & -0.7 \\[0.08cm]
$Q_s(4^+_1)$ & -0.41(12) & -0.7 \\[0.08cm]
$Q_s(2^+_2)$ & +0.1(1) & +0.6 \\[0.08cm]
$B(E2:2_1^+ \rightarrow 0_1^+)$ & 1252(117) & 1416 \\
\noalign{\smallskip}\hline
\end{tabular}
\caption{Same as Table \ref{tab:values128} but for $^{130}$Xe. Experimental data are taken from \cite{Morrison20a,Angeli13a,Stone20a,Huang21a,Wang21a}.}
\label{tab:values130}
\end{table}

\subsubsection{Average deformation}
\label{sec:avgdef130}

\begin{figure}[t!]
    \centering
    \includegraphics[width=0.70\linewidth]{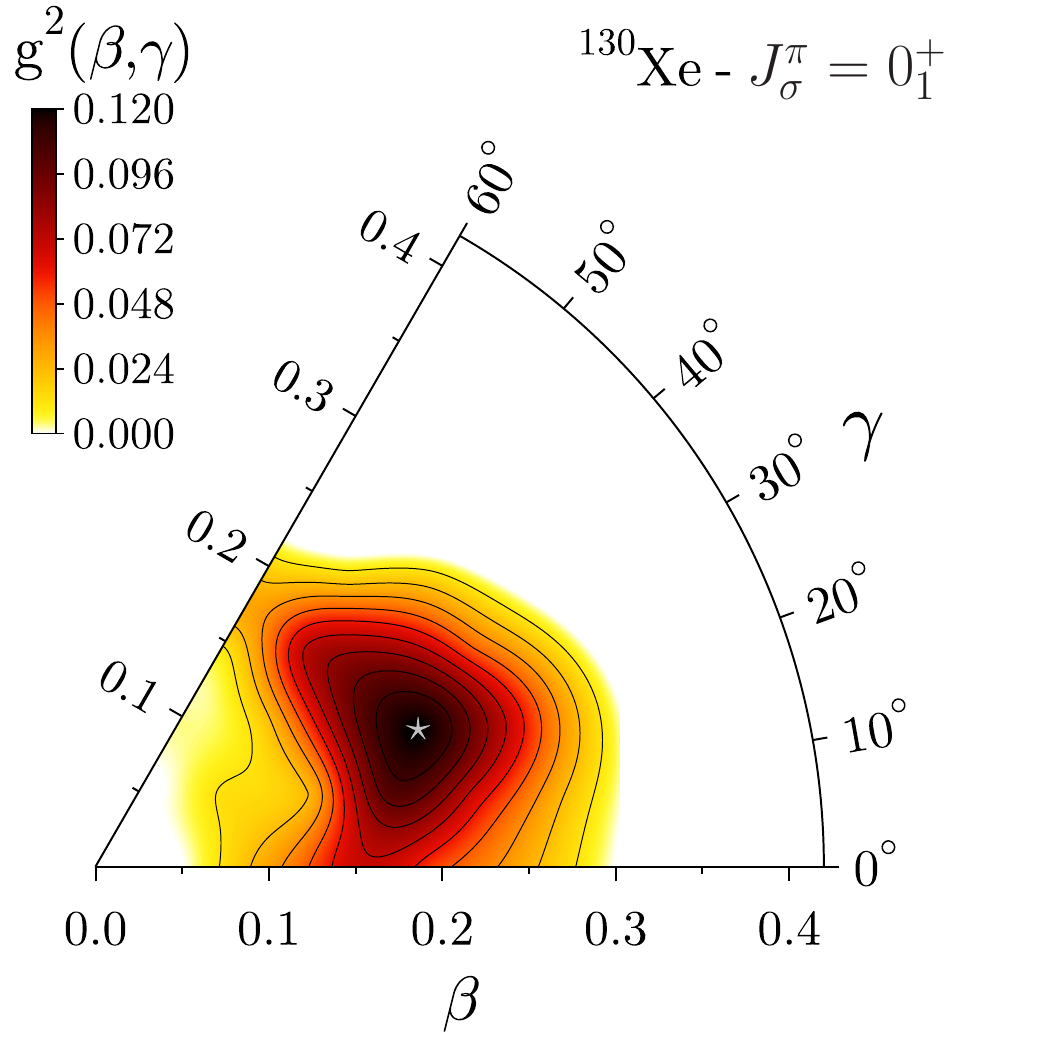}
    \caption{Same as Fig.~\ref{fig:xe128_coll} but for $^{130}$Xe.
    The maximum, indicated by a silver star, is located at a deformation of $\beta=0.20$ and $\gamma=23^\circ$.}
    \label{fig:xe130_coll}
\end{figure}

Finally, we plot in Fig.~\ref{fig:xe130_coll} the scwf for the ground state of $^{130}$Xe. Similarly to the two other isotopes discussed previously, the scwf is peaked at triaxial deformations. Actually, the maximum is located at a deformation of $\beta=0.20$ and $\gamma=23^\circ$ very similar to the one of $^{128}$Xe in spite of the energy surfaces being overall more rigid against $\beta$-deformation.

In Table \ref{tab:def130}, we report the deformation parameters obtained using the different methods presented in Sec.~\ref{sec:128}. Given the good agreement between the experimental and calculated $B(E2:2^+_1 \rightarrow 0^+_1)$ transition probabilities, we obtain a theoretical value
of $\beta_r (0_1^+) = 0.18$ for the rigid rotor elongation that is close to the experimental one of 0.17(1). By contrast, while the calculated moments for the $2^+_1$ and $2^+_2$ states have same size but opposite sign, which as previously discussed is compatible with the picture of an asymmetric rotor, it is less clear for the experimental moments. The moments have indeed opposite sign, but the absolute value of $Q_s(2^+_2)$ appears to be significantly smaller than the one of $Q_s(2^+_1)$. Their respective estimated error bars, however, are so large that it remains possible they are actually very close in value.

\begin{table}[t!]
\centering
\begin{tabular}{ccc}
\hline\noalign{\smallskip}
Quantity & Theory & Experiment \\
\noalign{\smallskip}\hline\noalign{\smallskip}
$\beta_r$ & 0.18 & 0.17(1) \\[0.08cm]
$\gamma_d$ & $21^\circ$ & $28^\circ$ \\
\noalign{\smallskip}\hline\noalign{\smallskip}
$\beta_{c}$ & 0.19 & \\[0.08cm]
$\Delta \beta_c$ & 0.04 &  \\[0.08cm]
$\gamma_{c}$ & $23^\circ$ &  \\[0.08cm]
$\Delta \gamma_c$ & $16^\circ$ &  \\
\noalign{\smallskip}\hline\noalign{\smallskip}
$\beta_k$ & 0.16 & 0.17(2) \\[0.08cm]
$\Delta \beta_k $ & 0.02 &  \\[0.08cm]
$\gamma_k$ & $28^\circ$ & $23(5)^\circ$  \\[0.08cm]
$\Delta \gamma_k $ & $12^\circ$ &  \\[0.08cm]
\noalign{\smallskip}\hline
\end{tabular}
\caption{Same as Table \ref{tab:def128} but for the $0_1^+$ ground state of $^{130}$Xe.
Experimental values are taken from Ref.~\cite{Morrison20a}.}
\label{tab:def130}
\end{table}

Repeating the analysis based on the Davydov's model already performed for $^{128}$Xe, we find here that with $E(2^+_1)+E(2^+_2) = 1.658$ MeV and $E(3^+_1) = 1.633$ MeV, the experimental energies nearly fulfils the equality $E(2^+_1)+E(2^+_2) = E(3^+_1)$ also for $^{130}$Xe. Also, the ratio $E(2^+_2)/E(2^+_1) = 2.09$ leads to an angle $\gamma_d (0_1^+)= 28^\circ$.
Concerning theoretical results, there is more of a discrepancy between the energies because $E(2^+_1)+E(2^+_2) = 1.740$ MeV while $E(3^+_1) = 1.953$ MeV, and from the ratio $E(2^+_2)/E(2^+_1) = 3.39$ we obtain the angle $\gamma_d (0_1^+)= 21^\circ$.

Using the scwf, we obtain the average deformations $\beta_{c}(0_1^+)=0.19$ and $\gamma_c (0_1^+)=  23^\circ$ that is consistent with the previous values.

Recently, the experimental values of the Kumar parameters for $^{130}$Xe were obtained from a low-energy Coulomb excitation experiment \cite{Morrison20a}. We report those values in Table \ref{tab:def130} alongside the ones coming from our MR calculations. There is a good agreement between the two as well as with the deformation parameters obtained from the other methods. Interestingly, and in contrast to $^{128}$Xe, the theoretical value of $\gamma_k (0_1^+)= 28^\circ$ is not at variance with the other estimates for the triaxial angle.

In conclusion, similarly to the two other xenon isotopes studied previously in this work, the structure of $^{130}$Xe is compatible with the picture of a triaxially deformed nucleus.


\subsection{Sequence of isotopes}
\label{sec:evol}

Having computed two successive even-even isotopes as well as the intermediate odd-mass one, we are in position to investigate if the experimental trends for relevant observables, such as the masses or the root-mean-square charge radius, are reproduced or not. To be consistent with experimental data, we consider here mostly the calculated $1/2^+_1$ state despite the fact it is not the ground state in our calculations.

\begin{figure}[t!]
    \centering
    \includegraphics[width=.90\linewidth]{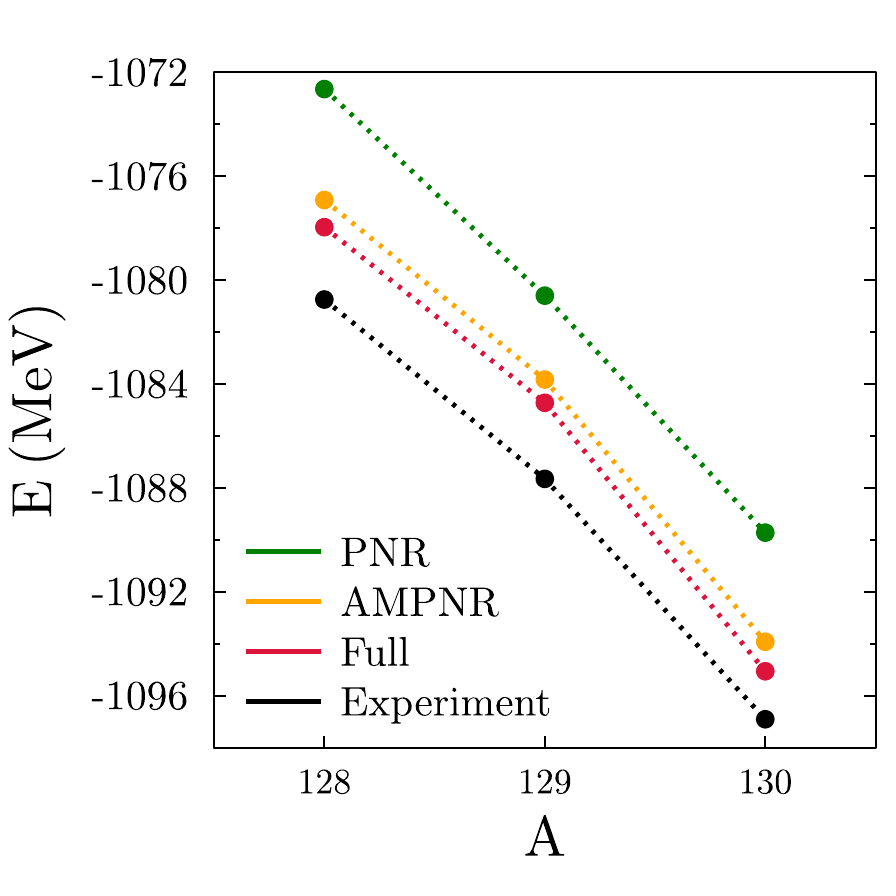}
    \caption{Binding energy as a function of the mass $A$ of the isotopes as indicated.}
    \label{fig:evol_ener}
\end{figure}

First of all, we are interested in the evolution of the binding energies as a function of the mass number and how it varies at different levels of approximation in our modeling. We show in Fig.~\ref{fig:evol_ener} the evolution for the experimental data as well as for three theoretical schemes: (i) the value computed  using the (interpolated) particle-number projected minima of positive parity for the three nuclei, (ii) the value computed using the  minima of the angular-momentum and particle-number restored (AMPNR) energy surfaces, and (iii) the value computed after the configuration mixing of projected Bogoliubov states. In cases (ii) and (iii), we take the binding energies of the $J^\pi_\sigma = 0^+_1$ level for $^{128,130}$Xe and of the $J^\pi_\sigma = 1/2^+_1$ level for $^{129}$Xe. We remark first that there is an important gain of about 4 MeV between the PNR and AMPNR curves, as discussed previously, and a much smaller gain between the AMPNR and the full calculations. That the configuration mixing usually brings less correlation energy than the restoration of symmetries is a well-known fact that has been observed before \cite{Bender06a}. We also notice that the full calculation exhibits a trend much closer to the experimental data than the simple PNR minima, although the decrease between $A=129$ and 130 is exaggerated in our calculations. We point out that when performing the analysis using the theoretical $9/2^-_1$ state instead (not shown here), the drop when going from $A=129$ to $A=130$ is less noticeable.

It is also interesting to compute the three-point mass difference 
\begin{equation}
\begin{split}
 \Delta^{(3)} (Z,N) = \frac{(-1)^N}{2} &\left[ E(Z,N+1) - 2 E(Z,N) \right. \\
                                     &\left. + E(Z,N-1) \right],
\end{split}
\end{equation}
that is widely used to characterize the amplitude for pairing correlations in nuclei, and to constrain the pairing strength of effective interactions during the parameter adjustment. In Table \ref{tab:delta3}, we compare this quantity at three levels of theoretical modeling with the experimental values of 1.17~MeV, obtained by considering the experimental $1/2^+_1$ ground state, and of 1.45~MeV, obtained by considering the $9/2^-_1$ excited state that is the ground state in our calculation. Analyzing first the case of the $1/2^+_1$ state, we see that the theoretical value for $\Delta^{(3)} (54,75)$ changes dramatically as when including correlations in the modeling: It starts from the value of 0.59~MeV at the PNR level that is too small by a factor of 2, gets a large increase to 1.59~MeV at the AMPNR level, and finally reaches 1.79~MeV when the full-fledged method is used. This is consistent with the trend of binding energies discussed above. By contrast, the three-point mass difference computed for the $9/2^-_1$ state does not change much with the level of approximation considered in our model and remains about 2 to 3 times smaller than the experimental value. While none of these values are in agreement with the experimental one, which is not so surprising given the fact that our calculations do not describe very well the energy spectrum of $^{129}$Xe, this limited analysis seems to indicate that the role of correlations in the computation of $\Delta^{(3)} (Z,N)$ depends on the state considered, and that they might be important in certain cases. A more detailed study on this aspect will certainly be needed. 
 
\begin{table}
\centering
\begin{tabular}{ccc}
\hline\noalign{\smallskip}
Description & $1/2^+_1$ & $9/2^-_1$ \\
\noalign{\smallskip}\hline\noalign{\smallskip}
Experiment & 1.17 & 1.45 \\[0.08cm]
PNR & 0.59 & 0.42 \\[0.08cm]
AMPNR &  1.59 & 0.62 \\[0.08cm]
Full  & 1.79 & 0.65\\[0.08cm]
\noalign{\smallskip}\hline
\end{tabular}
\caption{Values for the three-point formula $\Delta^{(3)}$ given in unit of MeV and considering either the $1/2^+_1$ or $9/2^-_1$ state of $^{129}$Xe.}
\label{tab:delta3}
\end{table}

\begin{figure}[t!]
    \centering
    \includegraphics[width=.90\linewidth]{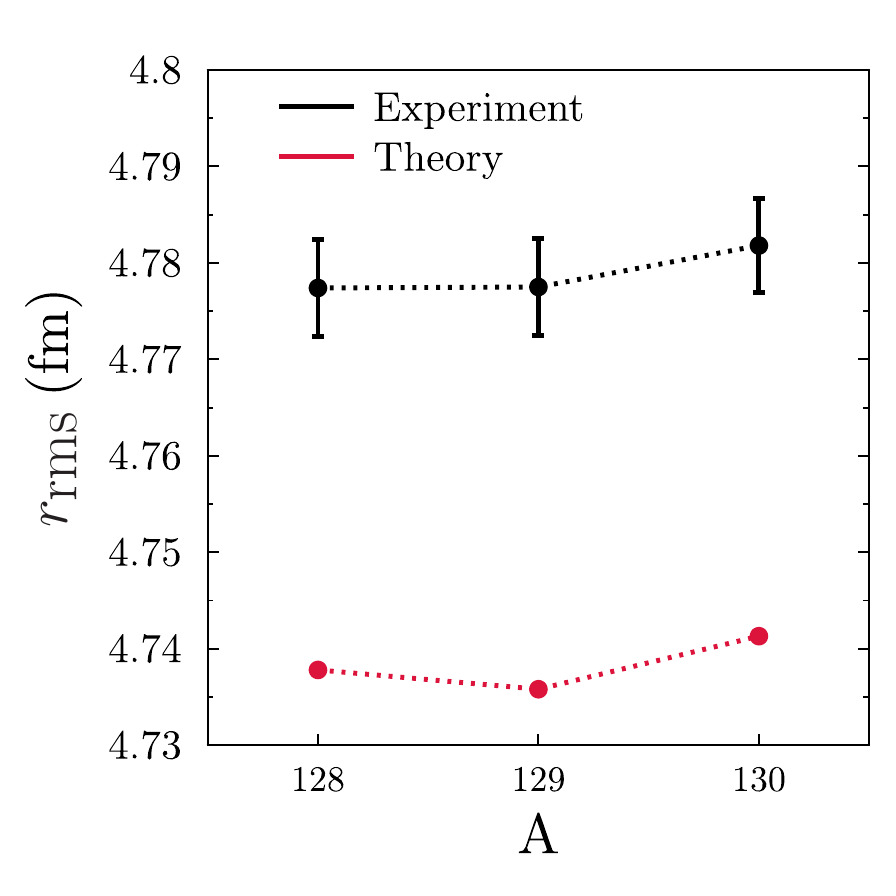}
    \caption{Root-mean-square charge radius as a function of the mass $A$ of the isotope. Experimental values are taken from Ref.~\cite{Angeli13a}.}
    \label{fig:evol_radii}
\end{figure}

Finally, we investigate the evolution of the root-mean-square charge radii as a function the mass number, which is shown in Fig.~\ref{fig:evol_radii}. Here, we consider directly the full MR-EDF calculations. All radii are somewhat too small, but in a consistent manner and still below the level of 1\% relative error. Given the experimental error bars, it is not possible to make a strong statement regarding the evolution of the radii but both in calculated results (based on the $1/2^+_1$ state) and the experimental data, one finds the usual odd-even staggering, i.e.\ that the radius of the intermediate odd nucleus is smaller than the average of the radii of the adjacent even-even nuclei, superposed on an overall increase of the charge radius with mass number. The necessary ingredients for the correct description of this phenomenon are presently a matter of debate \cite{Reinhard17a,Groote20a}
and it is therefore gratifying to see that it is reproduced by our MR scheme. We also remark that these observations are consistent with the results obtained in recent MR-EDF calculations based on the Gogny EDF dedicated to the magnesium isotopic chain \cite{Borrajo17a}. More systematic studies are of course needed to check if this is a systematic feature.

%
%
\section{Conclusions}
\label{sec:conclu}

In this article, we investigated the structure of the isotopes $^{128,129,130}$Xe by performing advanced MR-EDF calculations using the Skyrme-type pseudo-potential SLyMR1. We obtained a fair description of the even-mass isotopes $^{128,130}$Xe concerning their basic spectroscopic properties as well as the energies of their low-lying excited states and the $E2$ transitions between them. The relative order of the states in the odd-mass nucleus $^{129}$Xe, however, is not reproduced, in particular with a wrong assignment for the spin and parity of the ground state. 

In particular, our results are consistent with the interpretation of these three isotopes as being triaxial. This had already been concluded early on based on phenomenological models that were partially, if not completely, adjusted to the available information on the excitation spectra of these nuclei. More recent data obtained from a low-energy Coulomb excitation experiment \cite{Morrison20a} provided for the first time direct empirical evidence for the triaxial shape of $^{130}$Xe, and data from high-energy $^{129}$Xe + $^{129}$Xe collisions performed at the LHC \cite{ATLAS:2022dov} can for time being also only be consistently explained when assuming that this isotope has a triaxial shape~\cite{Bally22a}.

To further improve the quality of our description, it would be necessary, first and foremost, to improve the properties of the effective interaction used to model the strong interaction among nucleons. On the one hand, the parametrization SLyMR1 has the formal advantage that it corresponds to a proper Hamiltonian operator that can be safely used in multi-reference calculations without the possibility to encounter any of the unphysical behaviour discussed in Refs.~\cite{Lacroix09a,Bender09a,Duguet09a,Dobaczewski07a}. On the other hand, it possesses several deficiencies, two of which clearly compromise the description of the Xe isotopes studied here. One is its very small effective pairing strength that leads to very weak pairing correlations in most states. As a consequence, SLyMR1 might overly favour deformed states over spherical ones, and also might tend to reduce the coupling between states with very different deformation in a shape-mixing MR calculation, as pairing in general increases the overlaps and operator matrix elements between quasi-particle vacua that have a different sequence of occupied single-particle levels. The other deficiency is related to the predictions of SLyMR1 for the eigenvalues of the single-particle Hamiltonian, i.e.\ what is usually called the single-particle energies. These quantities are not directly observable, and their link to observable quantities depends on the correlations that are explicitly treated in the wave function \cite{Duguet15a}. The discussion of $^{129}$Xe illustrates that in a MR calculation the connection between single-particle levels and observed band heads becomes even more complicated as is already the case in self-consistent single-reference EDF calculations. In general, angular-momentum projection decomposes the blocked quasi-particle vacua into several states of same total angular-momentum \cite{Bally21a}, and these components projected from different quasi-particle vacua at same or other deformations are mixed at the MR level, with an overall energy gain that is different for each band head. As the dominant contributions to the full mixed states still can be traced back through the MR calculation, it can nevertheless be concluded that specific relative shifts of the positions of single-particle levels at sphericity on the order of a few several hundred keV could significantly improve the description of the low-lying excitation spectrum of $^{129}$Xe in a MR calculation that uses the same collective space, and in particular give a ground-state with correct angular momentum.\footnote{We recall that adding further correlations at the MR level, for example by including octupole-deformed reference states that are also parity projected in addition to the projections already considered here, can be expected to shift again the relative positions of the MR band heads when calculated with the same interaction, meaning that slightly different predictions for the single-particle energies would be needed to reach a similar agreement with data in that case.}
Deficiencies of the predictions for single-particle energies are not specific to SLyMR1, but rather a global problem of all existing EDFs~\cite{Bonneau07a} that requires further attention in the future.

There is also an interest to increase the variational space within which one generates the reference states for the MR calculation. The excitation energies of states at large angular momentum can most probably be improved upon by including additional reference states that are optimised for projecting out states with non-zero angular momenta by being constrained to finite average values of angular momentum with the help of a cranking constraint \cite{Borrajo15a}. Doing so can be expected to compress the spectrum as demonstrated in Ref.~\cite{Egido16b}. The $2_1^+$ and $4_1^+$ states of $^{128}$Xe and the $2_1^+$ level of $^{130}$Xe, however, are already a little bit too low in energy in calculations without cranking, hence the description of these low-lying states in even-mass isotopes would probably be slightly degraded. Unfortunately, adding a further collective coordinate to the configuration mixing comes at a large additional computational cost and was out of the scope for the present study. The same applies to adding reference states with two or three blocked quasi-particles, which might in particular be relevant for the description of the low-lying excited $0^+$ states in the even-even isotopes that do not exhibit shape coexistence \cite{Sun03a}.

While electric quadrupole moments are reasonably well described for most states in the three isotopes for which there are data, we often find quite large deviations between calculation and experiment for the magnetic moments. Part of those differences can be expected to have their origin in a mismatch of the actual configurations and a deficiency of the SLyMR1 parameterization to induce spin polarisation. For the reasons sketched in Sec.~\ref{sec:spectrum129}, however, there are also open questions concerning the very nature of a consistent $M1$ operator for nuclear EDF calculations that require further future study.

%
%
\begin{acknowledgements}
We would like to thank Jean-Yves Ollitrault for useful discussions and Vittorio Som{\`a} for proofreading the manuscript.
This project has received funding from the European Union’s Horizon 2020 research and innovation programme under the Marie Sk\l{}odowska-Curie grant agreement No.~839847. 
M.~B.~acknowledges support by the Agence Nationale de la Recherche, France, Grant No.~19-CE31-0015-01 (NEWFUN). G.G.~is funded by the Deutsche Forschungsgemeinschaft (DFG, German Research Foundation) under Germany's Excellence Strategy EXC2181/1-390900948 (the Heidelberg STRUCTURES Excellence Cluster), within the Collaborative Research Center SFB1225 (ISOQUANT, Project-ID 273811115).
The calculations were performed by using HPC resources from GENCI-TGCC, France (Contract No.~A0110513012).
\end{acknowledgements}
%
%
\begin{appendices}

\section{Expressions for the matrix elements of $\hat{E}^{(n)}_2$}
\label{sec:appred}

Let us first define the electric quadrupole operator $\hat{E}_2$ as the rank-2 tensor with components
\begin{equation}
  \hat{E}_{2\mu} = \sum_{i=1}^{A} \hat{e}_i \, \hat{r}_i^2 \, Y_{2\mu} (\hat{\theta}_i, \hat{\phi}_i), 
\end{equation}
where $\mu \in \llbracket -2, 2 \rrbracket$, $\hat{e}_i$ and ($\hat{r}_i, \hat{\theta}_i, \hat{\phi}_i$) are the electric charge and spherical coordinates of the $i$-th particle, respectively, and $Y_{2\mu}$ are the usual spherical harmonics. 

When considering the electric quadrupole moments of a classical ellipsoid, the finite sum of the above quantum mechanical operator is replaced by an integral over the volume of the ellipsoid and the charge operator by a charge density \cite{Kumar72a}.

In the present work, we consider for the products $\hat{E}^{(n)}_2$ the standard expressions given in the literature \cite{Cline86a,Srebrny06a} 
\begin{align}
 \hat{E}^{(2)}_2 &= \left[ \hat{E}_2 \times \hat{E}_2 \right]_0 , \\
 \hat{E}^{(3)}_2 &= \left[ \left[ \hat{E}_2 \times \hat{E}_2 \right]_2 \times \hat{E}_2 \right]_0 , \\
 \hat{E}^{(4)}_2 &= \left[ \hat{E}^{(2)}_2 \times \hat{E}^{(2)}_2 \right]_0 = \hat{E}^{(2)}_2 \hat{E}^{(2)}_2 , \\
 \hat{E}^{(5)}_2 &= \left[ \hat{E}^{(2)}_2 \times \hat{E}^{(3)}_2 \right]_0 =\hat{E}^{(2)}_2 \hat{E}^{(3)}_2 , \\
 \hat{E}^{(6)}_2 &= \left[ \hat{E}^{(3)}_2 \times \hat{E}^{(3)}_2 \right]_0 = \hat{E}^{(3)}_2 \hat{E}^{(3)}_2 .
\end{align}
With these definitions, by inserting the identity operator $\sum_{\Gamma,\sigma} \ket{\Psi^\Gamma_\sigma}\bra{\Psi^\Gamma_\sigma}$ between subparts of the operators coupled to zero angular momentum as well as using the expressions of matrix elements of irreducible tensor products \cite{Varshalovich88a}, we obtain
\begin{align}
 \label{eq:e1}
 \elma{\Psi^\Gamma_\sigma}{\hat{E}^{(2)}_2}{\Psi^\Gamma_\sigma} &= \frac{1}{\sqrt{5}(2J+1)} \sum_{J_1 \sigma_1} (-1)^{-J+J_1} \nonumber \\
   &\phantom{=} \times \elma{\Psi^{\Gamma}_{\sigma}}{|\hat{E}_2|}{\Psi^{\Gamma_1}_{\sigma_1}} \elma{\Psi^{\Gamma_1}_{\sigma_1}}{|\hat{E}_2|}{\Psi^{\Gamma}_{\sigma}}  , \\
 \elma{\Psi^\Gamma_\sigma}{\hat{E}^{(3)}_2}{\Psi^\Gamma_\sigma} &=  \frac{(-1)^{2J}}{(2J+1)}  \sum_{J_1 \sigma_1} \sum_{J_2 \sigma_2} 
   \left\{ \begin{array}{ccc} 2 & 2 & 2 \\ J_2 & J & J_1 \end{array}\right\}  \nonumber \\
   &\phantom{=} \times \elma{\Psi^{\Gamma}_{\sigma}}{|\hat{E}_2|}{\Psi^{\Gamma_1}_{\sigma_1}} \elma{\Psi^{\Gamma_1}_{\sigma_1}}{|\hat{E}_2|}{\Psi^{\Gamma_2}_{\sigma_2}} \nonumber \\
   &\phantom{=} \times \elma{\Psi^{\Gamma_2}_{\sigma_2}}{|\hat{E}_2|}{\Psi^{\Gamma}_{\sigma}} ,   \\
  \elma{\Psi^\Gamma_\sigma}{\hat{E}^{(4)}_2}{\Psi^\Gamma_\sigma} &= \sum_{\sigma_1} | \elma{\Psi^\Gamma_\sigma}{\hat{E}^{(2)}_2}{\Psi^\Gamma_{\sigma_1}}|^2 ,\\
  \elma{\Psi^\Gamma_\sigma}{\hat{E}^{(5)}_2}{\Psi^\Gamma_\sigma} &= \sum_{\sigma_1}  \elma{\Psi^\Gamma_\sigma}{\hat{E}^{(2)}_2}{\Psi^\Gamma_{\sigma_1}}
      \elma{\Psi^\Gamma_{\sigma_1}}{\hat{E}^{(3)}_2}{\Psi^\Gamma_{\sigma}} , \\
  \elma{\Psi^\Gamma_\sigma}{\hat{E}^{(6)}_2}{\Psi^\Gamma_\sigma} &= \sum_{\sigma_1} | \elma{\Psi^\Gamma_\sigma}{\hat{E}^{(3)}_2}{\Psi^\Gamma_{\sigma_1}} |^2 ,
  \label{eq:e6}
  \end{align}
where we have used the shorthand notation $\Gamma_i \equiv \linebreak[4] (J_i MNZ\pi)$ for the set of quantum numbers that differ only by their angular momentum 
and the fact that the eletric quadrupole operator does neither change the parity nor the number of particles of the system such that $\sum_{\Gamma_i} \rightarrow \sum_{J_i}$.
The expressions \eqref{eq:e1}-\eqref{eq:e6} are of course equalities only when summing over all possible states with relevant combinations of quantum numbers. In practice, however,
one is limited to a finite set of states such that the Kumar deformation parameters determined from these sums have some truncation error.

\end{appendices}
%
%
\interlinepenalty=10000
\bibliography{biblio.bib}
%
%
\end{document}